\documentclass[aps,pre,twocolumn,showpacs,superscriptaddress,groupedaddress,floatfix]{revtex4}

\usepackage{graphicx}   % needed for figures
\usepackage{amsmath}    % ERIK: don't use "eqnarray" but "align" instead!
\usepackage{hyperref}

\begin{document}

\widetext
\title{Finite-size scaling investigation of the liquid-liquid critical point in ST2 water
and its stability with respect to crystallization}

\author{T. A. Kesselring}
\affiliation{Computational Physics, IfB, ETH Zurich, Schafmattstrasse 6, 8093 Zurich, Switzerland}

\author{E. Lascaris}
\affiliation{Center for Polymer Studies and Department of Physics, Boston University, Boston, MA 02215}

\author{G. Franzese}
\affiliation{Departament de F\'{\i}sica Fonamental, Universitat de Barcelona, Diagonal 645, 08028 Barcelona, Spain}

\author{S. V. Buldyrev}
\affiliation{Department of Physics, Yeshiva University, 500 West 185th Street, New York, NY 10033}

\author{H. J. Herrmann}
\affiliation{Computational Physics, IfB, ETH Zurich, Schafmattstrasse 6, 8093 Zurich, Switzerland}
\affiliation{Departamento de F\'{\i}sica, Universidade Federal do Cear\'{a}, Campus do Pici, 60451-970 Fortaleza, Cear\'{a}, Brazil}

\author{H. E. Stanley}
\affiliation{Center for Polymer Studies and Department of Physics, Boston University, Boston, MA 02215}

\date{\today}

\begin{abstract}

The liquid-liquid critical point scenario of water hypothesizes the
existence of two metastable liquid phases---low-density liquid (LDL) and high-density
liquid (HDL)---deep within the supercooled region.  
The hypothesis originates from computer simulations of the ST2 water
model, but the stability of the LDL phase with respect to the crystal
is still being debated. 
We simulate supercooled ST2 water at constant pressure, constant
temperature and constant number of molecules $N$ for $N\leq 729$ and
times up to 1~$\mu$s. We observe clear differences between the two liquids, both
structural and dynamical.  Using several methods, including finite-size
scaling, we confirm the presence of a liquid-liquid phase transition
ending in a critical point. 
We find that the LDL  is stable with respect to the crystal in 98\% of
our runs (we perform 372 runs for LDL or LDL-like states), and in 100\% of our
runs for the two largest system sizes ($N=512$ and 729, for which we perform 136
runs for LDL or LDL-like states).
In all these runs tiny crystallites grow and then melt within 1~$\mu$s. 
Only for $N\leq 343$ we observe six events (over 236 runs for LDL or LDL-like
states) of spontaneous crystallization after crystallites reach an estimated
critical size of about $70 \pm 10$ molecules. 

\end{abstract}


\maketitle

%%%%%%%%%%%%%%%%%%%%%%%%%%%%%%%%%%%%%%%%%%%%%%%%%%%%%%%%%%%%%%%%%%%%%%%%%%%%%%%

\section{Introduction}

For many centuries, water and its anomalies have been of much interest to scientists.
A particular rise of interest occurred in the late 1970s after experiments done by
Angell and Speedy seemed to imply some kind of critical phenomenon in supercooled
liquid water at very low temperatures \cite{Angell1973,Speedy1976,Angell1982,Speedy1982}.
Even though liquid water experiments are limited by spontaneous crystallization
below the homogenous nucleation temperature ($T_H \approx 233$~K at 1~bar),
it is possible to further explore the phase diagram by quenching water to far
lower temperatures \cite{Burton1935,Bruggeller1980,Mishima1984}.
The result of these experiments is an amorphous solid, i.e. a glassy ice, 
corresponding to an out-of equilibrium state that is very stable with
respect to the equilibrium crystalline ice phase. The amorphous
depends on the applied pressure: at low pressure, below $\approx 0.2$~GPa,
the low density amorphous ice (LDA) is formed, while at higher
pressure the high density amorphous ice (HDA) is observed  \cite{Loerting2006}.
It has been shown by Mishima {\it et al.} that these two amorphous
ices are separated by a reversible abrupt change in density that
resembles in all its respects an equilibrium 
first order phase transition \cite{Mishima1985,Mishima1994,Mishima1998a,Mishima1998b}.

Raising the temperature of either LDA or HDA does not turn the sample into a
liquid, but leads once again to spontaneous crystallization (around $T_X \approx 150$~K).
In fact, between $T_H$ and $T_X$, often called the ``no man's land''
of bulk water, crystallization occurs at a time scale that is too short for 
current experimental methods, although a new technique is possibly
succeeding in the task of measuring the metastable liquid phase \cite{Nilsson2012pc}.
Computer simulations of water, however, involve time scales small enough to witness
spontaneous crystallization and are therefore able to explore liquid
water in the  ``no man's land''.
In 1992 Poole {\it et al.} \cite{Poole1992} performed a series of molecular dynamics simulations
using the ST2 water model \cite{Stillinger1974}, using the reaction
field method for the long-range interactions (ST2-RF), 
and discovered a liquid-liquid
phase transition ending in a critical point, separating a low density liquid (LDL)
and a high density liquid (HDL).
These two liquids can be considered to be the liquid counterparts of
the LDA and HDA, respectively.

The existence of the critical point also allows one to understand X-ray spectroscopy
results \cite{Tokushima2008,Huang2009,Nilsson2011,Wikfeldt2011},
explains the increasing correlation length in bulk water upon cooling
as found experimentally \cite{Huang2010}, 
the hysteresis effects \cite{Zhang2011} and the dynamic behavior of
protein hydration water \cite{Mazza2011,Franzese2011,Bianco2012a}.
It would be consistent with a range of thermodynamical and dynamical anomalies 
\cite{Kumar2011, Sciortino1990,
  Starr1999a,Kumar2008a,Kumar2008b,Franzese2009,delosSantos2011,delosSantos2012,Mazza2012} 
and experiments \cite{Franzese2008,Stanley2009,Stanley2010,Stanley2011}.

Many more computer simulations investigating the phenomenology of the liquid-liquid critical
point (LLCP) have been performed since then
\cite{Harrington1997,Franzese2001,Franzese2003,Franzese2007a,Hsu2008,Stanley2008,
Oliveira2008,Mazza2009,Franzese2010, Stokely2010,Corradini2010a,Vilaseca2010,Vilaseca2011,
Xu2011,Gallo2012,Strekalova2012b,Strekalova2012c,Bianco2012b}.
Detailed studies using ST2-RF have been made by Poole {\it et al.}
\cite{Poole2005} 
using molecular dynamics, while  
Liu {\it et al.} simulated ST2 with Ewald summation (ST2-Ew) for  the electrostatic 
long-range potential using Monte Carlo \cite{Liu2009,Liu2010}.
Also in other water models the liquid-liquid phase transition (LLPT)
and its LLCP are believed to be found, 
for example by Yamada {\it et al.} in the TIP5P model \cite{Yamada2002},
by Paschek {\it et al.} in the TIP4P-Ew model \cite{Paschek2008},
and in TIP4P/2005 by Abascal and Vega \cite{Abascal2010,Abascal2011}.

Recently Limmer and Chandler used Monte Carlo umbrella sampling
to investigate the ST2-Ew model, but claimed to have found only one
liquid metastable phase (HDL)
rather than two \cite{Limmer2011}.
They therefore concluded that LDL does not exist because it is unstable with
respect to either the crystal or the HDL phase. The emphasis in their
work is about the difference between a metastable phase,
i.e. separated from the stable phase by a finite free-energy barrier,
and an unstable state, where the free-energy barrier is absent and the
state does not belong to a different phase.

Shortly after, Poole {\it et al.} \cite{Poole2011} and Kesselring 
{\it et al.} \cite{Kesselring2012} presented results using standard
molecular dynamics for ST2-RF showing the occurrence of the LLCP with
both HDL and  LDL phases  metastable
with respect to the crystal, but with the LDL not unstable with
respect to either the crystal or the HDL.
This result was confirmed, 
using the same method as Limmer and Chandler, by
Sciortino {\it et al.} \cite{Sciortino2011} and Poole {\it et al.}
\cite{Poole2013} in ST2-RF water and
by Liu {\it el al.} in ST2-Ew water \cite{Liu2012}.

\begin{figure}[htb]
    \centering
	\includegraphics[angle=0,width=0.45\textwidth]{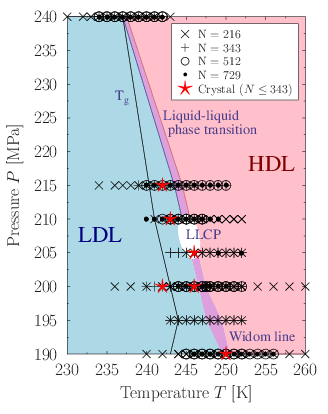}
    \caption{\label{FIG:phase_diagram}	
Overview of the state points at which simulations
have been performed, with the symbols indicating different
system sizes.  At high temperatures we observe a high-density
liquid state (HDL, shaded in pink), while at lower temperatures
we find a low-density liquid (LDL, in blue).  These
are separated by a region where the system is continuously
flipping between the two states, as seen in Fig.~\ref{FIG:rho_vs_time_for_different_T}.
This transition region (in purple) is identified as the liquid-liquid phase
transition line (LLPT) at high pressures, and the Widom
line at low pressures.  These lines join at the liquid-liquid
critical point (LLCP) estimated at $P_C = 208 \pm 3$~MPa and
$T_C=246 \pm 1$~K (see Sec.~\ref{SEC:locating_LLCP}).
At low temperatures the LDL (or LDL-like) region is bound by the glass
transition temperature $T_g$, 
below which we can no longer fully equilibrate the system within
100~ns, and consider the liquid to have become a glass (see Sec.~\ref{SEC:tau}).
For small sizes ($N \leq 343$) we observe spontaneous crystallization
within 1~ns-long simulations at six state points (indicated
by the red stars), all of them within the LDL (or LDL-like) region.  We
never observe crystallization for sizes $N = 512$ and 729 for
simulations of comparable duration. Because the probability
of crystalization should increase with $N$, this results suggest
that our cystallization events are a finite-size effect that
becomes negligeble for large sizes.  Crystallization events are
discussed in Sec~\ref{SEC:crystal_growth_and_melting}.
	}
\end{figure}

The aim of this paper is to confirm the presence of a liquid-liquid critical
point in water in the thermodynamic limit
using finite size scaling techniques, and confirm that LDL is a 
{\it bona fide} metastable liquid.
We use the ST2-RF model because it has been well-studied in the supercooled region,
making it easier to compare and verify our data.
In the supercooled phase it has a relatively large self-diffusion
compared to other water models, therefore 
suffers less from the slowing down of the dynamics at extremely low 
temperatures.
We explore a large region of the phase diagram of supercooled liquid ST2-RF water
(Fig.~\ref{FIG:phase_diagram}) using molecular dynamics simulations
with four different system sizes 
by keeping constant the number $N$ of molecules, the  
pressure $P$ and the temperature $T$ ($NPT$ ensemble).

\begin{figure}[htb]
    \centering
	\includegraphics[angle=0,width=0.45\textwidth]{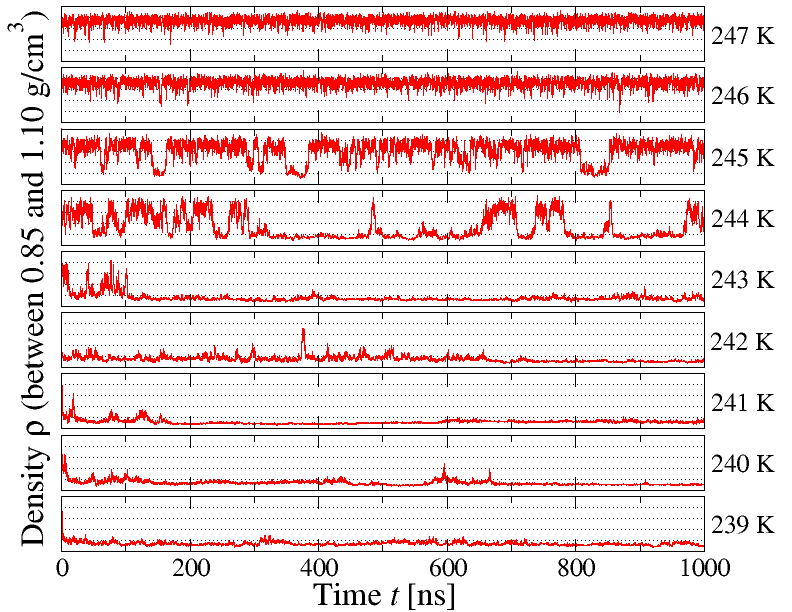}
    \caption{\label{FIG:rho_vs_time_for_different_T}
Phase flipping near the phase transition line ($P=215$~MPa, with $N=343$~molecules).
At high $T$ the system is in the HDL phase (with a density $\rho \simeq 1.03$~g/cm$^3$), while
at low $T$ the system is in the LDL phase (density $\rho \simeq 0.88$~g/cm$^3$).
However, near the phase transition line (at  $T\simeq 244.5$~K for
this pressure) the system is flipping between the two phases. 
    }
\end{figure}

Within the explored region we find both LDL and HDL, separated at high pressures
by a LLPT,
ending in a LLCP estimated at $P_C\approx 208$~MPa and $T_C\approx 246$~K.
This phase transition is particularly clear in Fig.~\ref{FIG:rho_vs_time_for_different_T}
where one can see from the density how the system continuously flips between the two states.
However, due to finite size effects this phase flipping also occurs below the critical
point along the Widom line (the locus of correlation length maxima) \cite{Xu2005, Franzese2007b}.
For this reason it is necessary to apply finite size scaling methods to establish the
exact location of the critical point.

For six state points and for small system size $N\leq 343$ we observe,
in only one over the (on average) seven simulations we performed for each state point,
irreversible crystal growth, indicated as red stars in Fig.~\ref{FIG:phase_diagram}.
Each of these crystallization events occurred within the LDL (or
LDL-like) region.
Analysis of these crystals revealed them to have a diamond cubic crystal structure.
As we will discuss later, because these events disappears for larger
systems, we ascribe these crystallization to finite-size effects.

We start in Sec.~\ref{SEC:simulation_details} with a description of the model
and the procedures that were used.  In Sec.~\ref{SEC:Sk} we discuss the use of the
intermediate scattering function to analyze the structure of the
liquid, and in Sec.~\ref{SEC:tau} 
its use in defining the correlation time.
The analysis of the liquid structure is continued in Sec.~\ref{SEC:structural_parameters} where
we define and compare a selection of structural parameters.
The parameter $d_3$ is found to be particularly well-suited to distinguish between the liquid and the crystal state,
and this fact is subsequently used in Sec.~\ref{SEC:crystal_growth_and_melting} where we discuss the
growth and melting of crystals within the LDL liquid.
In Sec.~\ref{SEC:locating_LLCP}, by defining the appropriate order parameter,
we show that the LLCP in ST2-RF belongs
to the same universality class as the 3D Ising model. We
accurately determine where the LLCP is located in the phase diagram
in the thermodynamic limit by applying finite size
scaling on the Challa-Landau-Binder parameter.
We discuss our results and present our conclusions in Sec.~\ref{SEC:conclusions}.

%%%%%%%%%%%%%%%%%%%%%%%%%%%%%%%%%%%%%%%%%%%%%%%%%%%%%%%%%%%%%%%%%%%%%%%%%%%%%%%

\section{Simulation details
\label{SEC:simulation_details}}

In the ST2 model \cite{Stillinger1974} each water molecule is represented by a rigid tetrahedral structure of five particles.
The central particle carries no charge and represents the oxygen atom of water.  It interacts with
all other oxygen atoms via a Lennard-Jones (LJ) potential,
$U_{\rm{LJ}}(r_{ij}) \equiv 4 \varepsilon \left[ (\sigma / r_{ij})^{12} - (\sigma / r_{ij})^6 \right]$
with $\varepsilon \equiv 0.31694$~kJ/mol and $\sigma \equiv 3.10$~\r{A}.
Two of the outer particles represent the hydrogen atoms.  Each of them carries a charge of $+0.2357$~e,
and is located a distance 1~\r{A} away from the central oxygen atom.
The two remaining particles carry a negative charge of $-0.2357$~e,
are positioned 0.8~\r{A} from the oxygen, and represent the lone pairs of a water molecule.

The electrostatic potential in ST2 is treated in a special way.
To prevent charges $a$ and $b$ from overlapping, the Coulomb potential
is reduced to zero at small distances:
\begin{align}
	U_{\rm{el}}(r_{ab}) \equiv S(r_{ij}) \frac{1}{4 \pi \epsilon_0} \frac{q_a q_b}{r_{ab}}
\end{align}
where $S(r_{ij})$ is a function that smoothly changes from one to zero as the distance between the molecules decreases,
\begin{align}
    S(r_{ij}) \equiv \left\{
		\begin{array}{ll}
        0                                                     &  (         r_{ij} \leq R_L) \\
		\frac{(r_{ij}-R_L)^2(3R_U-R_L-2r_{ij})}{(R_U-R_L)^3}  &  (R_L \leq r_{ij} \leq R_U) \\
        1 													  &           (r_{ij} \geq R_U) \\
		\end{array}
	\right.
\end{align}
with $R_L \equiv 2.0160$~\r{A}, $R_U \equiv 3.1287$~\r{A}, and where $r_{ij}$ is the
distance between the oxygen atoms of the interacting molecules.
In the original model a simple cutoff was used for the electrostatic interactions.
In this paper, however, we apply the reaction field method \cite{Steinhauser1982}
which changes the ST2 Coulomb potential to
\begin{align}
	U_{\rm{el}}(r_{ab}) \equiv S(r_{ij}) T(r_{ij}) \frac{q_a q_b}{4 \pi \epsilon_0} \left( \frac{1}{r_{ab}} + \frac{r_{ab}^2}{2R_{c}^{3}}  \right)
\end{align}
where $T(r_{ij})$ is another smoothing function:
\begin{align}
    T(r_{ij}) \equiv \left\{
		\begin{array}{ll}
        1                                                         &  (         r_{ij} \leq R_T) \\
		1 - \frac{(r_{ij}-R_T)^2(3R_c-R_T-2r_{ij})}{(R_c-R_T)^3}  &  (R_T \leq r_{ij} \leq R_c) \\
        0 							     						  &           (r_{ij} \geq R_c). \\
		\end{array}
	\right.
\end{align}
We use a reaction field cutoff $R_c \equiv 7.8$~\r{A} together with $R_T \equiv 0.95 R_c$.
These parameters define our ST2-RF water model and are the same that
were used in previous ST2-RF simulations.

For the LJ interaction we use a simple cutoff at the same distance of 7.8~\r{A}.
We do not adjust the pressure to correct for the effects of the LJ cutoff \cite{Horn2004,Allen1987}, since
these adjustments come from mean field calculations which become increasingly
weak as one approaches a critical point.

We use the SHAKE algorithm \cite{Ryckaert1977} to keep the relative position
of each particle within a ST2 molecule fixed.  The temperature and pressure are
held constant using a Nos\'{e}-Hoover thermostat \cite{Nose1984,Allen1987,Nose1991} together
with a Berendsen barostat \cite{Berendsen1984}.
In all simulations periodic boundary conditions are applied.

\begin{figure}[htb]
    \centering
	\includegraphics[angle=0,width=0.45\textwidth]{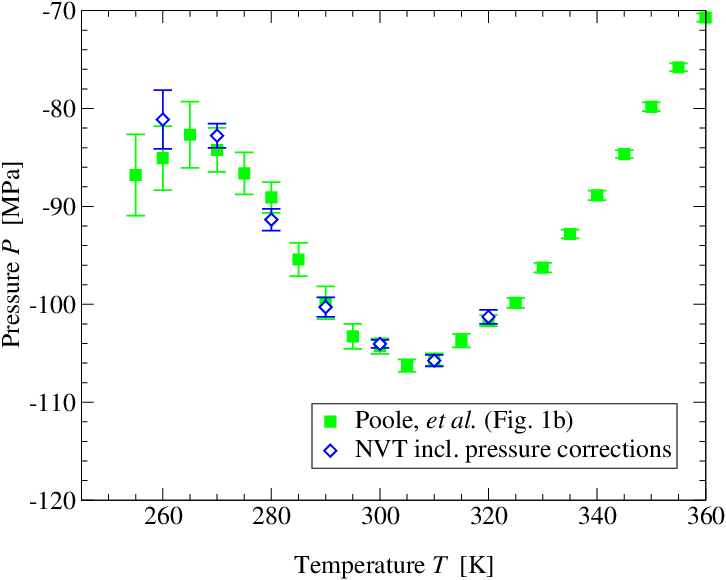}
    \caption{\label{FIG:comparing_with_Poole}
To validate our code, we compare our simulation results with those from Poole {\it et al.}~\cite{Poole2005}
at density $\rho=0.83$~g/cm$^3$ and for $N=216$ molecules.  We performed simualtions in the $NVT$ 
ensemble applying pressure corrections and find the same results as Ref.~\cite{Poole2005} within the error bars.
At this density the pressure correction due to the LJ cutoff (proportional to $\rho^2$) is equal to $-12.66$~MPa.
The variation of $P$ with $T$ along this isochore shows the occurence of both
a density maximum at 300~K and a density minimum near 265~K,
as at these state points $(\partial\rho / \partial T)_P = -\rho K_T (\partial P/\partial T)_V=0$
with $K_T>0$ the isothermal compressibility.
    }
\end{figure}

Our code is validated by simulating the same state points as
those published by Poole {\it et al.}, see Fig.~1b in
\cite{Poole2005}, where pressure corrections for the LJ cutoff were
applied in the $NVT$ (constant $N$, $T$ and volume $V$)
ensemble.
Averaging at each state point over 10 simulations with different initial conditions allows us to
estimate the error bars.
In Fig.~\ref{FIG:comparing_with_Poole} we compare our results for $N=216$ molecules and density 0.83~g/cm$^3$,
and find that our data, after pressure correction, matches that of  Ref.~\cite{Poole2005} well.

For each of the simulations done in the $NPT$ ensemble, we use the following protocol.
We first create a box of $N$ molecules at $n$ different initial
densities (with $n$ up to 21) ranging from 0.85 to 1.05~g/cm$^3$.
We then perform a 1~ns $NVT$ simulation at $T = 300$~K.  In this way we obtain $n$ independent
configurations all at $T = 300$~K in the prefixed range of densities.
Next, we use these independent configurations as starting points for $NPT$
simulations at $T = 265$~K and different pressures ranging from 190
to 240~MPa, and continue the simulation for an additional 1~ns.  This
results in $n$ independent configurations at $T = 265$~K and the given
pressure.
For all pressures considered here, this will lead the system into the HDL phase.
Finally the system is quenched to the desired temperature at the given
pressure, followed by 100--200~ns of equilibration time.
In Sec.~\ref{SEC:tau} it will be shown that this provides
enough time for the system to reach equilibrium for the state points
above the line marked with the label $T_g$ in Fig.~\ref{FIG:phase_diagram}

%%%%%%%%%%%%%%%%%%%%%%%%%%%%%%%%%%%%%%%%%%%%%%%%%%%%%%%%%%%%%%%%%%%%%%%%%%%%%%%

\section{Intermediate scattering function
\label{SEC:Sk}}

The intermediate scattering function $S(\mathbf{k},t)$ plays an essential role in the analysis of liquid structure,
since it is frequently measured in experiments as well as easily calculated from simulation data.
It describes the time evolution of the spatial correlation at the wave vector $\mathbf{k}$, and can be used
to distinguish between phases of different structure, such as LDL and
HDL or crystal.
It is defined as
\begin{align*}
S(\mathbf{k},t)   & \equiv
    \frac{1}{N} \left<
        \sum_{\ell,m}^N e^{i \mathbf{k} \cdot \left[ \mathbf{r}_{\ell}(t') - \mathbf{r}_{m}(t'+t) \right] }
    \right> _{t'}
\end{align*}
where $\left<  ... \right> _{t'}$ denotes averaging over simulation time $t'$,
and $\mathbf{r}_{\ell}(t')$ the position of particle $\ell$ at time $t'$.
For simplicity we only apply the intermediate scattering function to the oxygen atoms, which we denote as $S_{OO}(k,t)$.

Since the system has periodic boundary conditions, the components of $\mathbf{k}$ have discrete values $2 \pi n/L$,
where $L$ is the length of the simulation box and $n=1,2,3,\dots$.
We define $S_{OO}(k,t) \equiv \left< S_{OO}(\mathbf{k},t) \right>_n$ where the average is taken
over all vectors $\mathbf{k}$ with magnitude $k$ belonging to the $n$th spherical bin
$\pi (n-\tfrac{1}{2})/L \leq k < \pi (n+\tfrac{1}{2})/L$ for $n=2,3,\dots,300$.
Similarly, we define the structure factor $S_{OO}(k) \equiv \left< S_{OO}(k,t) \right>_t$ as the time-averaged
intermediate scattering function, with (unless indicated otherwise) the average taken over the whole duration of the run.

\begin{figure}[htb]
    \centering
	\includegraphics[angle=0,width=0.45\textwidth]{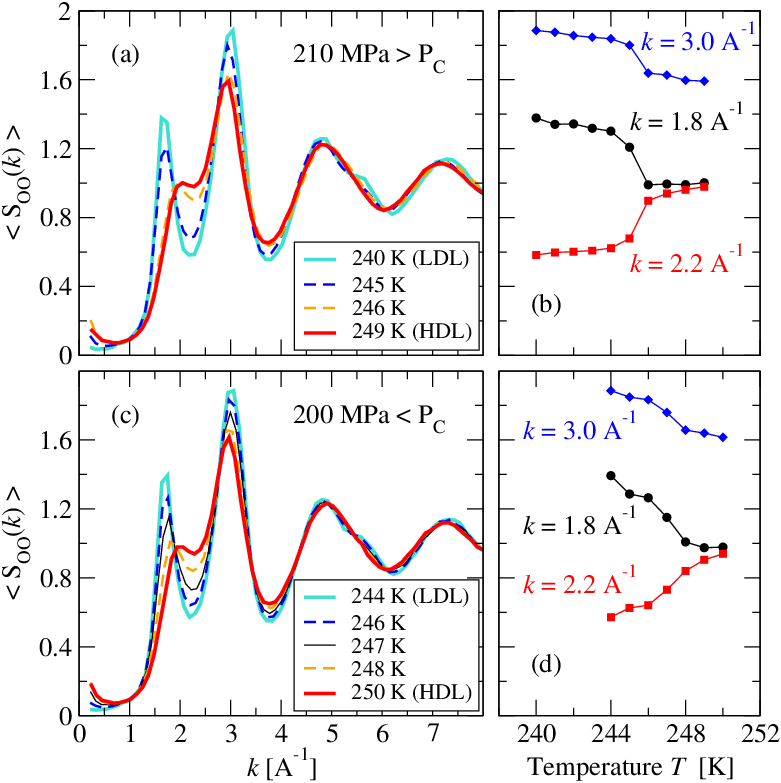}
    \caption{\label{FIG:Sk_vs_T}
The structure factor $S_{OO}(k)$ for a range of temperatures at (a) 210~MPa and (c) 200~MPa
for $N=729$.
(a) For $P > P_C$ the structure has a large change between $T = 245$
and 246~K, corresponding to the LDL-HDL first-order phase
transition.
(b) The value of $S_{OO}$ for $k$ corresponding to the
first maximum, the first minimum and the second maximum
as a function of $T$ for $P = 210$~MPa as in panel~(a).
(c) For $P < P_C$ the structure changes in a way that is smoother than
the case in panel~(a), with the more evident change occurring
between $T = 249$ and 250~K, corresponding to the crossing
of the Widom line, as marked by the value of $S_{OO}$ at first
maxima and minima in panel~(d).
    }
\end{figure}

We study $S_{OO}(k)$ above and below our estimate for the LLCP pressure.
At $P= 210$~MPa $> P_C$
(Fig.~\ref{FIG:Sk_vs_T}a,b)  
we observe a discontinuous change in the first two peaks of
$S_{OO}(k)$ as $T$ changes between 245 and 246~K, 
and a continuous change above and below this
temperatures. This is the expected behavior for a first order phase
transition occurring 
at $245$~K$\lesssim T\lesssim 246$~K and  $P= 210$~MPa between two
phases with different structure, consistent with our results in
Fig.~\ref{FIG:phase_diagram}. The fact that for both phases
$S_{OO}(k)\sim O(1)$ for all $k$ shows that both phases are
fluid. Indeed, for a crystal-like configuration, with a long-range
order, there would be at least one wave vector such that 
$S_{OO}(k)\sim O(N)$  \cite{Franzese2002}. Furthermore, the fact that at lower $T$ the
first peak increases and the other peaks only have minor changes indicates
that the lower-$T$ liquid has a smaller density than the higher-$T$
liquid. Therefore, this result show a first-order  phase transition
between the LDL at lower-$T$  and HDL at higher-$T$. 
This transition occurs at the same temperature at which 
we observe the phase flipping in density
(Fig.~\ref{FIG:rho_vs_time_for_different_T}) and corresponds to the
purple region at $P>P_C$ in Fig.~\ref{FIG:phase_diagram}.

\begin{figure}[htb]
    \centering
    \includegraphics[angle=0,width=0.45\textwidth]{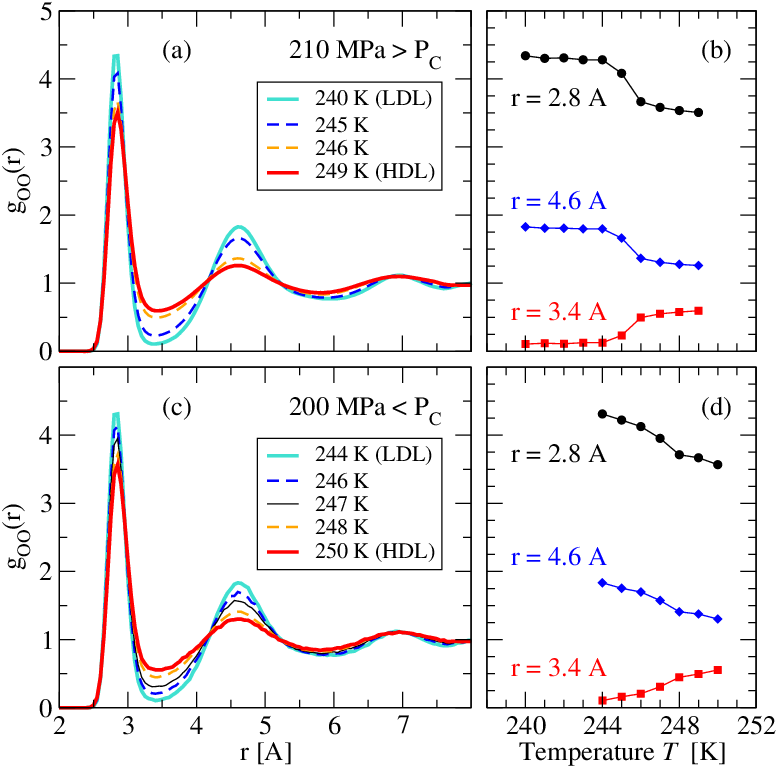}
    \caption{\label{FIG:rdf_vs_T}
The radial distribution function $g_{OO}(r)$ for the state points in
Fig.~\ref{FIG:Sk_vs_T}.
(a) For $P>P_C$ the main structural change between LDL and HDL is visible around the second
coordination shell at $r \simeq 4.6$~\AA\ and is stronger when $T$ changes between
245 and 246~K, as emphasized by the change of values of $g_{OO}$
for $r$ corresponding to the first maximum and minimum and
the second maximum in panel~(b).
(c) The transition from LDL to HDL is smoother for $P < P_C$ when the system is
crossing the Widom line, as shown by the variation of the
values of $g_{OO}$ in panel~(d).
    }
\end{figure}

The fact that the peaks of $S_{OO}(k)$ are sharper in LDL than HDL is
an indication that the LDL phase is more structured.
We can also observe that the major structural
changes in $S_{OO}(k)$ between  LDL  and HDL are for $k\simeq
1.8$ and 2.8~\AA$^{-1}$, corresponding to $r=4\pi/k\simeq 7$ and 4.5~\AA,
respectively, i.e. are for the third and the second  neighbor water
molecules. This change in the structure is consistent with a marked
shift inwards of the second shell 
of water with increased density, and almost no change in the first
shell (at $k\simeq 4.6$~\AA$^{-1}$ and $r\simeq 2.75$~\AA), as seen in
structural experimental data for 
supercooled heavy water interpreted with Reverse Monte Carlo method
\cite{Soper2000}. This changes are visible also in the OO radial
distribution function $g_{OO}(r)$ (Fig.~\ref{FIG:rdf_vs_T}a,b).

For $P<P_C$ (Fig.~\ref{FIG:Sk_vs_T}c,d)  by
increasing $T$ we observe that the first peak of  $S_{OO}(k)$ 
merges with the second, transforming continuously in a shoulder.
Same qualitative behavior is observed for $g_{OO}(r)$
(Fig.~\ref{FIG:rdf_vs_T}c,d). These quantities
show us also that the lower-$T$ structure is LDL-like, while the
higher-$T$ structure is HDL-like. However, the absence of any
discontinuous change in the structure implies the absence of a
first-order phase transition in the structure of the liquid. This is
consistent with the occurrence of a LLCP at the end of the first-order
phase transition somewhere between 200 and 210~MPa, at a temperature
between 245 and 250~K.
In Sec.~\ref{SEC:locating_LLCP} we shall apply a different method 
to locate the LLCP with more precision.

At $P<P_C$, in the one-phase region, we expect to find the Widom line
emanating from the LLCP. The Widom line is by definition the locus of maxima of
the correlation length, therefore, for general thermodynamic considerations
\cite{Franzese2007b} near the LLCP it must be also the locus of
maxima of the response functions. In particular, it must be the locus
where the isobaric heat capacity $C_P\equiv T(\partial S/\partial T)_P$,
where $S$ is the entropy of the system,
has its maximum along a constant-$P$ path. This maximum occurs where
the entropy variation with $T$ is maximum, expected where the
structural variation of the liquid is maximum, i.e. where the
derivatives of the values of $S_{OO}(k)$
(Fig.~\ref{FIG:Sk_vs_T}d) and $g_{OO}(r)$
(Fig.~\ref{FIG:rdf_vs_T}d) with $T$ are
maximum. The interval of temperatures for each $P$ where this occurs
corresponds to the
purple region at $P<P_C$ in Fig.~\ref{FIG:phase_diagram}, 
indicated as the Widom line.

\begin{figure}[htb]
    \centering
    \includegraphics[angle=0,width=0.45\textwidth]{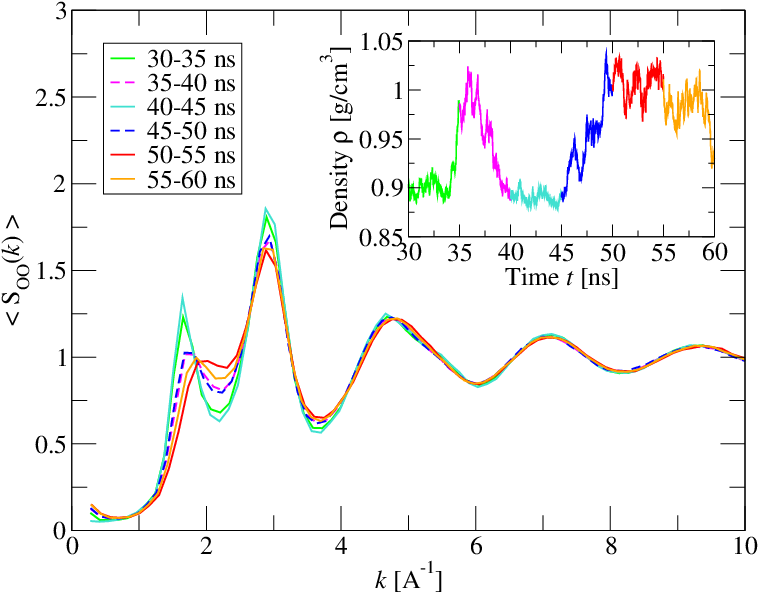}
    \caption{\label{FIG:Sk_vs_density}
As the density changes from $\rho(\rm{LDL})$ to $\rho(\rm{HDL})$, also the structure changes.
The inset shows how the density is changing with time for six consecutive time intervals of 10~ns,
with the corresponding $S_{OO}(k)$ shown in the main plot ($N=343$ at 200~MPa and 248~K).
    }
\end{figure}

It is actually possible to follow the structural changes during the simulation.
An example is given in Fig.~\ref{FIG:Sk_vs_density} where we focus on a 30~ns time period
of a simulation at 200~MPa and 248~K.  We divide this time period into six 5~ns intervals
and for each interval we calculate the intermediate scattering
function, time-averaged over those 5~ns. 
We observe that the liquid is LDL-like for the first and third
interval, having low density and  LDL-like $S_{OO}(k)$
 (first peak near 2~\r{A}$^{-1}$, separated from the second).
On the contrary, for the fifth and sixth interval the density is high
and  $S_{OO}(k)$ is HDL-like (the first peak
is merely a shoulder of the second peak), indicating that the liquid
is HDL-like.
For the second and fourth interval, the liquid has an intermediate
values of density and  $S_{OO}(k)$, indicating that it is a mix of
LDL-like and HDL-like structures.

%%%%%%%%%%%%%%%%%%%%%%%%%%%%%%%%%%%%%%%%%%%%%%%%%%%%%%%%%%%%%%%%%%%%%%%%%%%%%%%

\section{Correlation time
\label{SEC:tau}}

Apart from its use in structure analysis, the intermediate scattering function $S_{OO}(k,t)$
can also be used to define a correlation time $\tau$, i.e. the time it takes for a
system to lose most of its memory about its initial configuration \cite{Starr1999b,Kumar2006}.

\begin{figure}[htb]
    \centering
    \includegraphics[angle=0,width=0.45\textwidth]{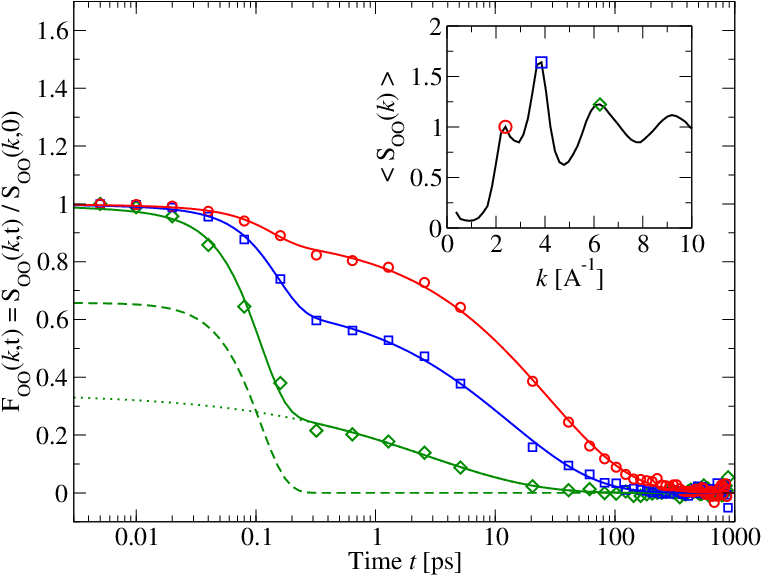}
    \caption{\label{FIG:def_tau_Foo}
Decay of $S_{OO}(k,t)$ with time, for $P=210$~MPa, $T=250$~K and $N=343$.
Symbols indicate $F_{OO}(k_i,t)$
% \equiv S_{OO}(k_i,t)/S_{OO}(k_i,0)$
for three different values of $k$: the first maximum of $S_{OO}(k)$ at $k_1$ (red circles), the
second maximum at $k_2$ (blue squares), and the third maximum $k_3$ (green diamonds).
Solid lines are fits according to Eq.~(\ref{EQ:tau_fit}).
%, which shows that $F_{OO}$ consists of a two-step process.  
The two components of  Eq.~(\ref{EQ:tau_fit}) are explicitly shown for $F_{OO}(k_3,t)$:
the green dashed line represents the $\beta$-relaxation and is given by
$[1 - A(k)] \exp[-(t/\tau_{\beta})^2]$, the green dotted line represents the $\alpha$-relaxation
and satisfies $A(k) \exp[-(t/\tau_{\alpha})^b]$.
The solid green line going through $F_{OO}(k_3,t)$ is the sum of both.
    }
\end{figure}

In Fig.~\ref{FIG:def_tau_Foo} we show how $S_{OO}(k,t)$ decays with time for a fixed value of $k$.
Its decay is characterized by two relaxation times, the $\alpha$-relaxation time $\tau_{\alpha}$ and the $\beta$-relaxation time $\tau_{\beta}$.
On very short time scales, the molecules do not move around much and each molecule is essentially stuck in a cage formed by its neighbors.
The $\beta$-relaxation time $\tau_{\beta}$ is of the order of
picoseconds.
On longer time scales, the molecule can escape from its cage and 
diffuse away from its initial position.
The time $\tau_{\alpha}$ is the relaxation time of this structural 
%diffusion 
process.

Mode-coupling theory of supercooled simple liquids predicts that \cite{Gallo1996}
\begin{align}
\label{EQ:tau_fit}
	F_{OO}(k,t) & \equiv S_{OO}(k,t)/S_{OO}(k,0) \notag \\
	         &      = [1 - A(k)] \, e^{-(t/\tau_{\beta})^2} + A(k) \, e^{-(t/\tau_{\alpha})^b}
\end{align}
The factor $A(k)$ is the Debye-Waller factor arising from the cage effect, which is independent of the temperature and follows
$A(k) = \exp(-a^2 k^2 / 3)$ with $a$ the radius of the cage.
We are able to fit Eq.~(\ref{EQ:tau_fit}) remarkably well to all our
data, as for example in Fig.~\ref{FIG:def_tau_Foo}.

\begin{figure}[htb]
    \centering
    \includegraphics[angle=0,width=0.45\textwidth]{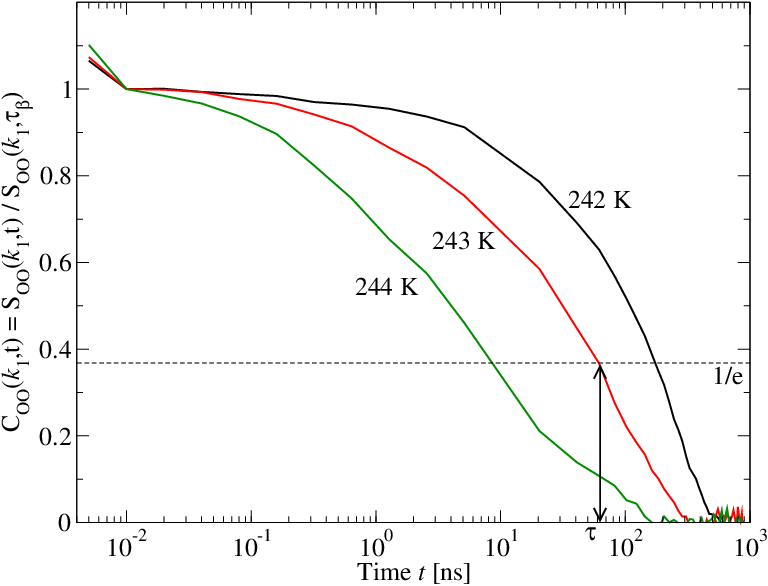}
    \caption{\label{FIG:def_tau_Coo}
Decay of $S_{OO}(k_1,t)$ with time, for three different temperatures at $P=210$~MPa ($N=343$).
Using $C_{OO}(k,t)$ 
%\equiv S_{OO}(k,t)/S_{OO}(k,\tau_{\beta})$ 
with $\tau_{\beta}$ as the approximately
time when $S_{OO}(k,t)$ reaches a plateau, it is possible to obtain a good estimate of $\tau_{\alpha}$.
Indicated here is the $\tau_{\alpha}$ for 243~K, equal to $\simeq 60$~ns.
At given $P$ and $T$  we define the correlation time $\tau$ as the
longest time $\tau_{\alpha}(k)$ for which $C_{OO}(k,\tau_{\alpha}) =
1/e$ 
(thin dashed black line).
    }
\end{figure}

Data in Fig.~\ref{FIG:def_tau_Foo} 
was collected every 10~fs for simulations of 1~ns.
This rate of sampling results in a large amounts of data and is
unfeasible for our runs up to 1000~ns. Therefore, for the 1000~ns runs
we collect data at 10~ps intervals. At this rate of sampling 
it is no longer possible to estimate $\tau_{\beta}$ or the cage size
$a$, but it is still possible to determine $\tau_{\alpha}$ accurately, utilizing
the fact that $S_{OO}(k,t)$ reaches a plateau near $t \approx \tau_{\beta}$.
One can therefore define 
\begin{equation}
C_{OO}(k,t) \equiv S_{OO}(k,t)/S_{OO}(k,\tau_{\beta}),
\end{equation}
which is $S(k,t)$ normalized by its value at the plateau (Fig.~\ref{FIG:def_tau_Coo}).
A good estimate of $\tau_{\alpha}$ is then the time for which $C_{OO}(k,\tau_{\alpha}) = 1/e \approx 0.37$.

From the shorter 1~ns runs (which were mostly done in the HDL regime) we find that the cage radius is $a = 0.35 \pm 0.09$ \AA\ with a 
stretching exponent of $b = 0.63 \pm 0.09$.
Both parameters $a$ and $b$ do not show a significant dependence on the state point within the studied range of temperature and pressure.

As shown in Fig.~\ref{FIG:def_tau_Foo}, different $k$ result in slightly different values for $\tau_{\alpha}$.
We use as the correlation time $\tau$ the largest value of $\tau_{\alpha}$ which is usually
found at $k=k_1$, the first maximum in $\left< S_{OO}(k) \right>$ (inset Fig.~\ref{FIG:def_tau_Foo}).

\begin{figure}[htb]
    \centering
	\includegraphics[angle=0,width=0.45\textwidth]{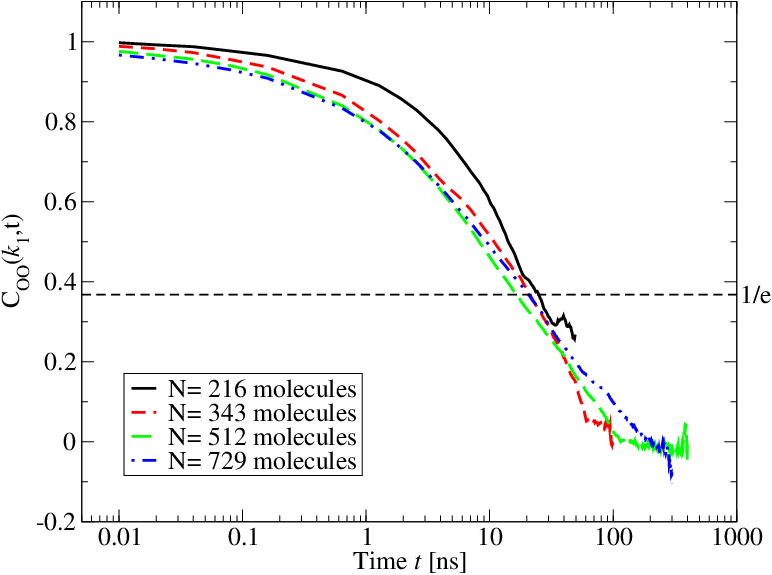}
    \caption{\label{FIG:tau_does_not_depend_on_N}
Correlation function for four systems with different sizes at 210~MPa and 243~K.
As the curves are all quite similar (which is a result we also find for the other state points),
we conclude that the system size has a negligible effect on the correlation time.
    }
\end{figure}

\begin{figure}[htb]
    \centering
    \includegraphics[angle=0,width=0.45\textwidth]{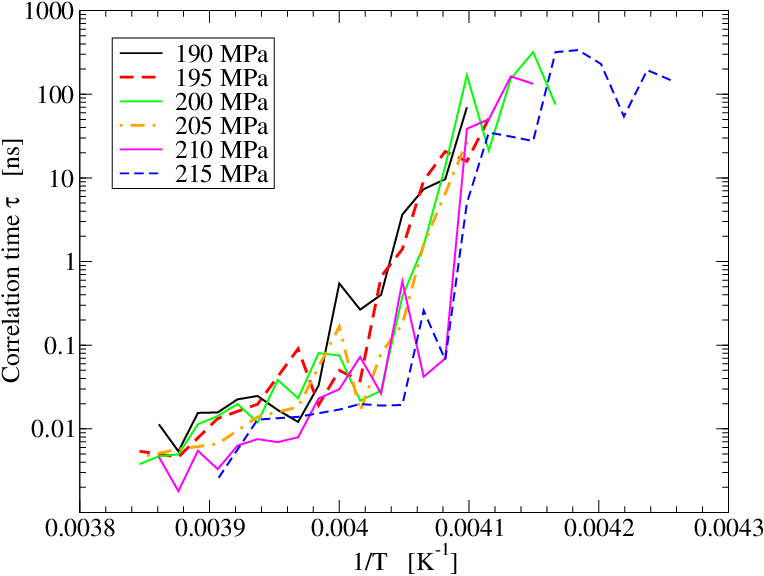}
    \caption{\label{FIG:tau_vs_P_and_T_for_N343}
Arrhenius plot of the correlation time $\tau$ for different pressures.
Errors on our estimates are of the order of the discontinuities along
the curves. 
At high temperatures (the HDL regime) the correlation
time is of the order of 10--100~ps, which jumps several orders up as we pass the phase transition line and enter the LDL regime.
To obtain this plot, we dismissed the simulations that had a significant increase
in $\tau$ because of crystal growth (Sec. \ref{SEC:crystal_growth_and_melting}).
    }
\end{figure}

As is to be expected, the correlation time does not seem to depend on the box size (Fig.~\ref{FIG:tau_does_not_depend_on_N}).
It does however depend strongly on the phase, which is evident from Fig.~\ref{FIG:tau_vs_P_and_T_for_N343}.

At high temperatures the system is in the HDL phase, and has a correlation time $\tau$ on the order of 10--100~ps.
As we decrease the temperature at fixed pressure, the value of $\tau$
has a large increase when we cross 
the phase transition line or the Widom line, depending if $P$ is above
or below $P_C$, respectively.
Apparently, the LDL states evolve nearly four orders of magnitude
slower than HDL states, with correlation times in the nanosecond range.

If we lower the temperature further, the correlation time slowly increases until the system becomes a glass rather than a liquid,
and we are no longer able to fully equilibrate the system.
As we can only run simulations up to 1000~ns, we consider the state points with a correlation time above 100~ns to be beyond our reach.
We therefore designate the effective glass transition temperature $T_g$
as the temperature for which $\tau > 100$~ns (see Fig.~\ref{FIG:phase_diagram}).

%%%%%%%%%%%%%%%%%%%%%%%%%%%%%%%%%%%%%%%%%%%%%%%%%%%%%%%%%%%%%%%%%%%%%%%%%%%%%%%

\section{Structural parameters
\label{SEC:structural_parameters}}

Apart from the intermediate scattering function, there are other ways to
quantify the structure of a liquid.  In this section we shall examine
several structural parameters, and determine which of those are the
most effective in distinguishing between LDL, HDL, and the 
%diamond-structured 
crystal.
For simplicity, we approximate the center of mass of a water molecule
with the center of its oxygen atom.

The structural parameters are designed to distinguish between different
phases by analyzing the geometrical structure.
This is typically done by evaluating the spherical harmonics $Y_{\ell}^{m}(\varphi,\vartheta)$
for a particular set of neighboring atoms, with $\varphi$ and $\vartheta$ the polar angles
between each pair of oxygen atoms in that set.
In this paper we consider two different sets: we define the first coordination shell $n_1(i)$
to be the four nearest neighbors of molecule $i$, and define the second coordination shell $n_2(i)$
as the fifth to sixteenth nearest neighbors
(the sixteenth nearest neighbors minus those in the first shell).

Different values of $\ell$ are sensitive to different symmetries.
The spherical harmonics with $\ell=3$, for example, are sensitive to a diamond structure.
Those with $\ell=6$ are more sensitive to the hexagonal closest packing (hcp) structure.
Since we expect the liquid and crystal structures to be hcp, diamond, or a mix of these, we focus primarily on $\ell=3$ and $\ell=6$.

%  %  %  %  %  %  %  %  %  %  %  %  %  %  %  %  %  %  %  %  %  %  %  %  %  %  %

\subsection{Parameters $q_3$ and $q_6$}

\begin{figure}[htb]
    \centering
\includegraphics[angle=0,width=0.45\textwidth]{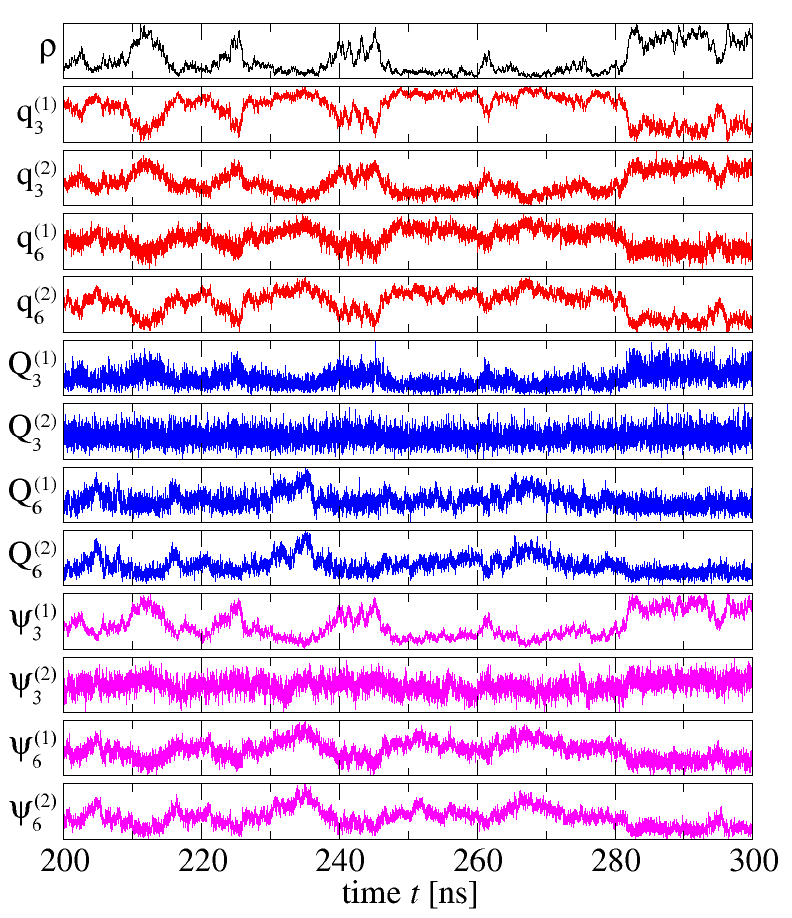}
    \caption{\label{FIG:Structural_parameters_vs_time}
Fluctuations of the density and the global structural parameters as a function of time.
The parameters are shown for one run using 343 molecules at 200~MPa and 248~K,
the same as in Fig.~\ref{FIG:Sk_vs_density}.
Parameters $q_{3}^{(1)}$, $q_{6}^{(2)}$, and $\psi_{3}^{(1)}$ (defined
in the text) are as sensitive as $\rho$ to the difference between
LDL-like and HDL-like structures, while the others are more
noisy, being $Q_{3}^{(2)}$ and $\psi_{3}^{(2)}$ much less sensitive than all the
others.  $Q_{6}^{(s)}$ and $\psi_{6}^{(s)}$, for both $s = 1$ and 2, have similar
behaviors that might be related to the temporary appearance
of crystal-like structures.
    }
\end{figure}

\begin{figure}[htb]
    \centering
	\includegraphics[angle=0,width=0.45\textwidth]{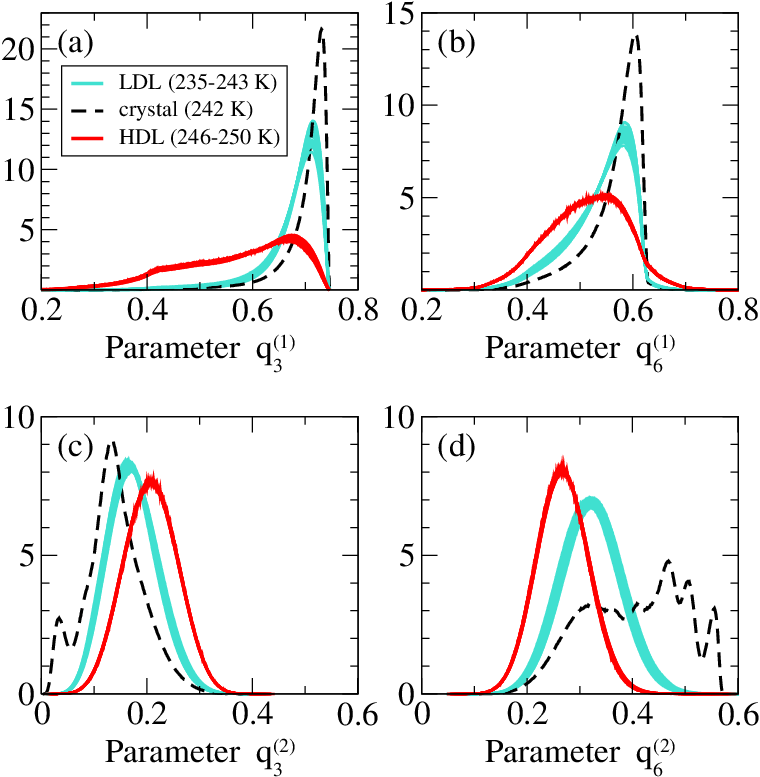}
    \caption{\label{FIG:histograms_q}
Histograms of $q_{\ell}^{(s)}$ for $\ell=3,6$ and coordination shells $s=1,2$ at 215~MPa with $N=343$ molecules.
The solid red (dark) curves correspond to HDL structures, and the
solid blue (light) curves to LDL structures.  The dashed
black curve corresponds to the crystal structure found in run C
described in Sec.~\ref{SEC:crystal_growth_and_melting}.
The parameter $q_{3}^{(1)}$ (a) discriminates
better between HDL and LDL structures, while the
parameter $q_{6}^{(2)}$ discriminates better between liquid-like and
crystal-like structures. Parameters in (b) and (c) are much
less sensitive to structural changes.
    }
\end{figure}

All parameters defined in this section are based on $q_{\ell,m}^{(s)}(i)$ which quantifies the local symmetry around molecule $i$.
It is defined as
\begin{align}
\label{EQ:def_qi}
  q_{\ell,m}^{(s)}(i) \equiv \frac{1}{N_s} \sum_{j \in n_{s}(i)} Y_{\ell}^{m}(\varphi_{ij},\vartheta_{ij}) \ \ \ -\ell \leq m \leq \ell
\end{align}
where $\ell$ and $m$ are integers,
 $s=1,2$ indicates the shell we are considering, with $N_s$ the number of molecules within that
shell (i.e. $N_1\equiv 4$ for the first coordination shell, and
$N_2\equiv 12$ for the second).
$Y_{\ell}^{m}$ is normalized according to $\int |Y_{\ell}^{m}|^2 \sin(\vartheta) \mathrm{d}\varphi \mathrm{d}\vartheta = 1$.
We can consider $q_{\ell,m}^{(s)}(i)$ as a vector $\mathbf{q}_{\ell}^{(s)}(i)$ in a $(4\ell+2)$-dimensional Euclidean space
having components Re$(q_{\ell,m}^{(s)}(i))$ and
Im$(q_{\ell,m}^{(s)}(i))$.
%with $m=-\ell,\dots,-1,0,1,\dots,\ell$.
This means that we can define an inner product
\begin{align}
\label{EQ:inner_prod}
  \mathbf{q}_{\ell}^{(s)}(i) \cdot \mathbf{q}_{\ell}^{(s)}(j)
                                \equiv \sum_{m=-\ell}^{\ell} \left[ \mathrm{Re}(q_{\ell,m}^{(s)}(i)) \, {\rm Re}(q_{\ell,m}^{(s)}(j))  \right.
  \mspace{20mu} \notag\\
					    				\left. + \, \mathrm{Im}(q_{\ell,m}^{(s)}(i)) \, \mathrm{Im}(q_{\ell,m}^{(s)}(j))  \right]
\end{align}
and a magnitude
\begin{align}
\label{EQ:magnitude}
  q_{\ell}^{(s)}(i)  \equiv
                     \sqrt{\mathbf{q}_{\ell}^{(s)}(i) \cdot \mathbf{q}_{\ell}^{(s)}(i)}.
\end{align}
The local parameter $q_{\ell}^{(s)}(i)$ is one way to distinguish between different structures,
and can be used to label individual molecules as LDL-like or HDL-like.
We can convert it into a global parameter by averaging over all molecules,
\begin{align}
  q_{\ell}^{(s)} \equiv \frac{1}{N} \sum_{i=1}^{N} q_{\ell}^{(s)}(i).
\end{align}
%
%where $N$ is the total number of molecules.
In Fig.~\ref{FIG:Structural_parameters_vs_time} we see that all global $q_{\ell}^{(s)}$ are sensitive to the difference between LDL and HDL,
especially $q_{3}^{(1)}$ and $q_{6}^{(2)}$.
We conclude that the structural difference is visible in both the first and second shell, and that LDL and HDL differ mostly in the amount of 
diamond structure of the first shell and the amount of hcp structure
in the second shell. 
This is confirmed by the histograms in Fig.~\ref{FIG:histograms_q}, in
which the largest difference between LDL and HDL is seen in
$q_{3}^{(1)}$ and, next, in $q_{6}^{(2)}$. The latter is the parameter
that better discriminate with respect to the crystal structure.

%  %  %  %  %  %  %  %  %  %  %  %  %  %  %  %  %  %  %  %  %  %  %  %  %  %  %

\subsection{Global parameters $Q_3$ and $Q_6$}

An alternative approach, as used by Steinhardt {\it et al.} \cite{Steinhardt1983}, is to
first average $q_{\ell,m}^{(s)}(i)$ over all molecules, defining
$Q_{\ell,m} \equiv \sum_{i=1}^N q_{\ell,m}^{(s)}(i)$, then calculate the magnitude
\begin{align}
  Q_{\ell}^{(s)} \equiv  \frac{1}{N} \left( \sum_{m=-\ell}^{\ell}  Q_{\ell,m} Q_{\ell,m}^{*} \right)^{1/2}.
%  Q_{\ell}^{(s)} \equiv \left( \frac{4 \pi}{2\ell + 1} \sum_{m=-\ell}^{\ell}  Q_{\ell,m} Q_{\ell,m}^{*} \right)^{1/2}
\end{align}

Our calculations show that the parameters  $Q_3^{(s)}$ and $Q_6^{(s)}$,
with $s=1$, 2, are not efficient in discriminating between LDL and HDL
(Fig.~\ref{FIG:Structural_parameters_vs_time}),
although $Q_6\equiv Q_6^{(1)}$ has been proposed recently as a good
parameter to this goal \cite{Limmer2011} and consequently has been used by several
authors \cite{Sciortino2011,Liu2012,Poole2013}. 
In particular, we observe that there is not much
correlation between the fluctuations of $Q_{\ell}^{(s)}$ 
and those of the density, except for $Q_{3}^{(1)}$.

\begin{figure}[htb]
    \centering
	\includegraphics[angle=0,width=0.45\textwidth]{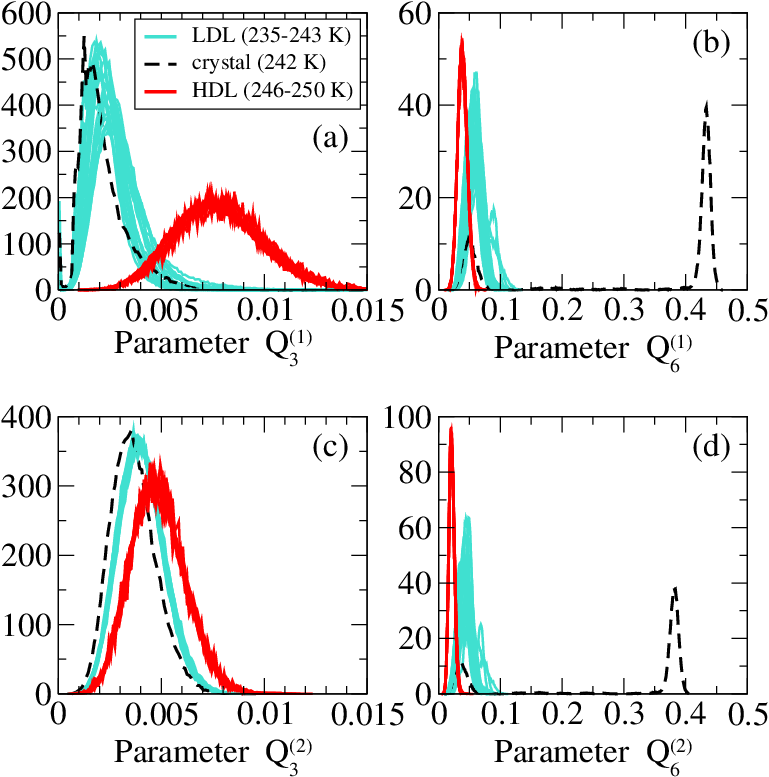}
    \caption{\label{FIG:histograms_Q}
Histograms of $Q_{\ell}^{(s)}$ for $\ell=3,6$ and coordination shells $s=1,2$ at 215~MPa with $N=343$ molecules.
The symbols are as in Fig.~\ref{FIG:histograms_q}.
The parameter $Q_{6}^{(s)}$, for the first
shell in (b) and the second in (d), shows a clear difference between
the liquid-like structures and the crystal-like structure,
but not between the two liquids.  Note that scales on x-axis
in panels (a) and (c) are one order of magnitude smaller than
those in panels (b) and (d).  As a consequence, $Q_{3}^{(s)}$, for the
first shell in (a) and the second in (c), is much less sensitive
to structural changes than $Q_{6}^{(s)}$.
    }
\end{figure}

However, we confirm
that $Q_6^{(1)}$ and $Q_6^{(2)}$ are excellent parameter to distinguish
between the liquids (LDL and HDL) and the crystal, being the value of
$Q_{6}^{(s)}$ approximately $10$ times larger for the crystal than it is for the liquids
 (Fig.~\ref{FIG:histograms_Q}). This large increase of $Q_{6}^{(s)}$
 for crystal-like structures might be related to the few instances in
 Fig.~\ref{FIG:Structural_parameters_vs_time} where an increase in
 $Q_{6}^{(s)}$ corresponds to a decrease of density (such as within
 interval $t=230$--237~ns), consistent with  the observation that the
 crystal-like structures have a density comparable to the
 LDL structure and smaller than the HDL structure.

\begin{figure}[htb]
    \centering
    \includegraphics[angle=0,width=0.45\textwidth]{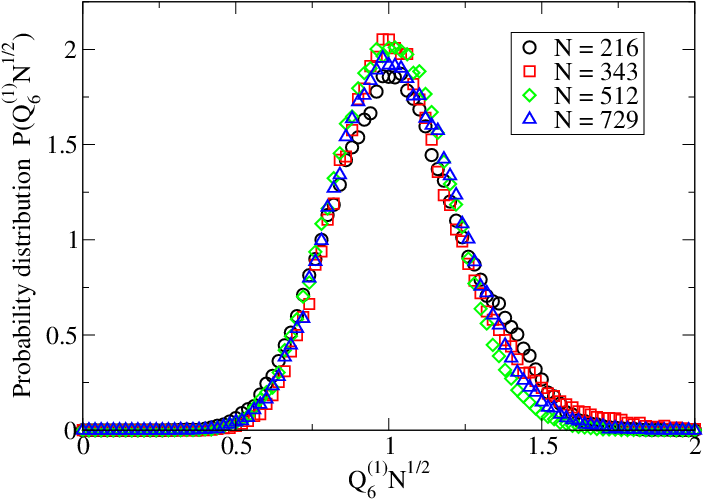}
    \caption{\label{FIG:Q6_vs_N}
Finite size scaling of parameter $Q_{6}$ in the LDL phase (210~MPa, 243~K).
The probability distribution function of $Q_{6} N^{1/2}$ is independent of the system size $N$,
which means LDL scales like a liquid in the thermodynamic limit: $Q_{6} \sim N^{-1/2}$.
    }
\end{figure}

To confirm that LDL remains a liquid in the thermodynamic limit, we
look at how $Q_{6}$ changes with the system size. 
For liquids $Q_{6}$ scales like $N^{-1/2}$ while for crystals the
value $Q_{6}$ remains finite as $N \rightarrow \infty$. We find that 
the probability distribution functions of $Q_{6} N^{1/2}$ for $N=216$, 343, 512, and 729
overlap, which means that $Q_{6} N^{1/2}$ is independent of
the system size, therefore $Q_{6} \sim N^{-1/2}$ (Fig.~\ref{FIG:Q6_vs_N}).
We conclude that the metastable LDL is not transforming into the
stable crystal in the thermodynamic limit. This implies that the LDL
and the crystal phase are separate by a free-energy barrier that is
higher than $k_BT$ at the temperatures we consider here and that the
system equilibrates to the stable (crystal) phase only in a time scale
that is infinite with respect to our simulation time (1000~ns), as occur
in experiments for metastable phases. Therefore, the LDL is a
{\it bona fide} metastable state. Our conclusion is consistent with
recent calculations by other authors
\cite{Sciortino2011,Liu2012,Poole2013}.

%  %  %  %  %  %  %  %  %  %  %  %  %  %  %  %  %  %  %  %  %  %  %  %  %  %  %

\subsection{Bond parameters $d_3$ and $\psi_3$}

We define the bond order parameter $d_{\ell}^{(s)}$ similar to that defined by
Ghiringhelli {\it et al.} in Ref.~\cite{Ghiringhelli2008}, where 
the quantity $d_3^{(1)}(i,j)$ characterizes the bond between molecules $i$ and $j$, and is designed
to distinguish between a fluid and a diamond structure.
The local parameter $d_{\ell}^{(s)}(i,j)$ is defined as the cosine of the angle between the
vectors $\mathbf{q}_{\ell}^{(s)}(i)$ and $\mathbf{q}_{\ell}^{(s)}(j)$:
\begin{align}
  d_{\ell}^{(s)}(i,j) \equiv \frac{ \mathbf{q}_{\ell}^{(s)}(i) \cdot \mathbf{q}_{\ell}^{(s)}(j) }
                                  { \left| \mathbf{q}_{\ell}^{(s)}(i) \right| \left| \mathbf{q}_{\ell}^{(s)}(j) \right|}
\end{align}
with the inner product and magnitude as defined in Eqs.~(\ref{EQ:inner_prod}) and (\ref{EQ:magnitude}).

A crystal with a perfect diamond structure has $d_3^{(1)}(i,j) = -1$ for all bonds.
For a graphite crystal only the bonds within the same layer (three out
of four) have $d_3^{(1)}(i,j) = -1$,
while the bonds connecting atoms in different layers (one out
of four) have $d_3^{(1)}(i,j) = -1/9$.

\begin{figure}[htb]
    \centering
    \includegraphics[angle=0,width=0.45\textwidth]{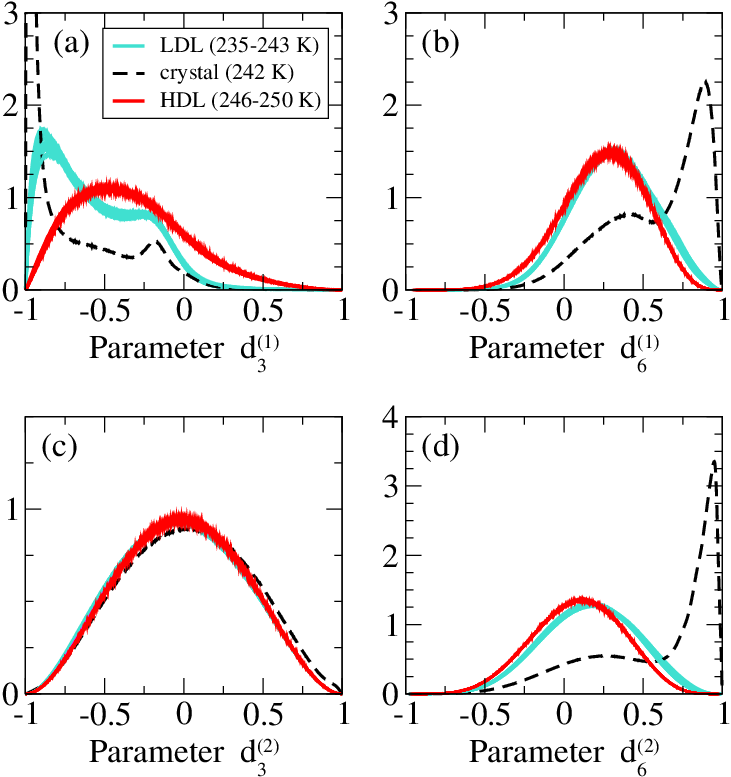}
    \caption{\label{FIG:histograms_d}
Histograms of $d_{\ell}^{(s)}$ for $\ell=3,6$ and coordination shells $s=1,2$ at 215~MPa with $N=343$ molecules.
The symbols are as in Fig.~\ref{FIG:histograms_q}.
Apart from $d_{3}^{(1)}$ in (a), these
parameters do not distinguish well between the two different
liquid-like structures, but $d_{3}^{(1)}$ and $d_{6}^{(s)}$
for the first shell (b) and the second (d) are suitable to distinguish between the
crystal and the liquids. The parameter $d_{3}^{(2)}$ in (c) is
remarkably the same for the three structures.
    }
\end{figure}

We find that the parameters $d_{\ell}^{(s)}$ for $\ell=3$, 6 and $s=1$, 2 
do not distinguish well between the two different liquid-like
structures, but that $d_{3}^{(1)}$ and $d_{6}^{(s)}$ for both $s=1$ and 2 
are suitable to discriminate between the crystal-like structure and the
liquids (Fig.~\ref{FIG:histograms_d}). In particular, 
for the crystal, most molecules have $d_3^{(1)} < -0.87$, and we
therefore consider a molecule to be part of a crystal if at least three out
of its four bonds with its nearest neighbors have $d_3^{(1)} < -0.87$.
This is the same cutoff used by Ghiringhelli {\it et al.} in \cite{Ghiringhelli2008}.

\begin{figure}[htb]
    \centering
	\includegraphics[angle=0,width=0.45\textwidth]{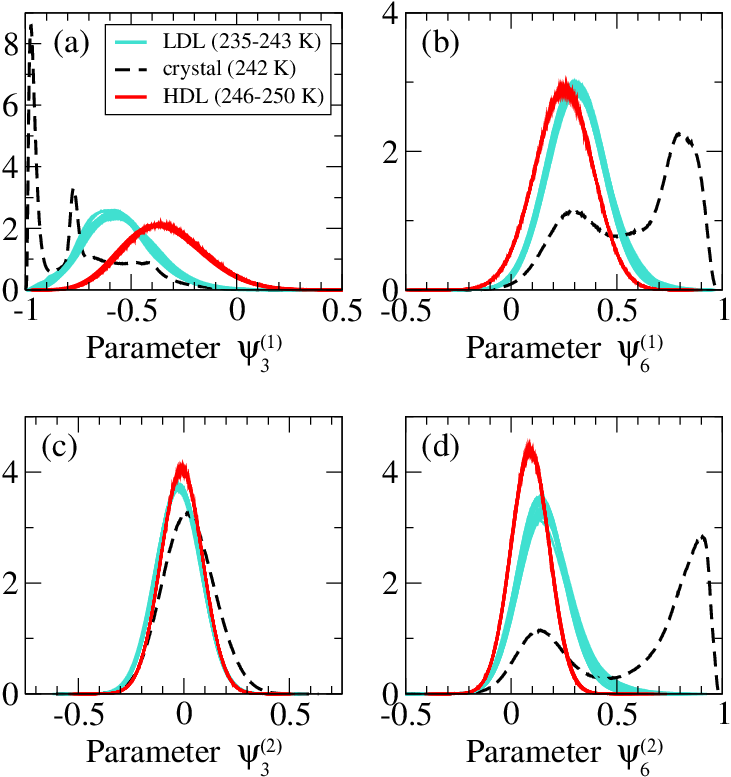}
    \caption{\label{FIG:histograms_psi}
Histograms of $\psi_{\ell}^{(s)}$ for $\ell=3,6$ and coordination shells $s=1,2$ at 215~MPa with $N=343$ molecules.
The symbols are as in Fig.~\ref{FIG:histograms_q}.
Each $\psi_{\ell}^{(s)}(i)$ has similar features as
the corresponding $d_{\ell}^{(s)}(i,j)$ in Fig.~\ref{FIG:histograms_d}.
    }
\end{figure}

The global parameter associated to $d_{\ell}^{(s)}(i,j)$ is defined as 
\begin{align}
  \psi_{\ell}^{(s)} \equiv \frac{1}{N} \sum_{i=1}^{N} \psi_{\ell}^{(s)}(i)
\end{align}
where
\begin{align}
\psi_{\ell}^{(s)}(i) \equiv \frac{1}{4}\sum_{j=1}^{4} d_{\ell}^{(s)}(i, j)
\end{align}
is the average of $d_{\ell}^{(s)}(i, j)$ over the first four nearest neighbors
of the molecule $i$. We observe that each $\psi_{\ell}^{(s)}(i)$ has the
same features of the corresponding $d_{\ell}^{(s)}(i, j)$, with 
$\psi_{3}^{(1)}(i)$ discriminating well between the crystal-like and the
liquids-like structures (Fig.~\ref{FIG:histograms_psi}).
We observe that $\psi_{3}^{(1)}$ discriminates well between LDL-like
and HDL-like structures
(Fig.~\ref{FIG:Structural_parameters_vs_time}), while $\psi_{6}^{(s)}$
for $s=1$ and 2 might be able to emphasize the temporary appearance of 
crystal-like structures, as noted for $Q_6^{(s)}$.

%%%%%%%%%%%%%%%%%%%%%%%%%%%%%%%%%%%%%%%%%%%%%%%%%%%%%%%%%%%%%%%%%%%%%%%%%%%%%%%

\section{Growth and melting of crystal nuclei
\label{SEC:crystal_growth_and_melting}}

In a small percentage of our simulations, the system was found to spontaneously crystallize.
These are interesting events because spontaneous crystallization of water in molecular dynamics is extremely rare;
only recently Matsumoto {\it et al.} were the first to successfully simulate the freezing of water on a computer \cite{Matsumoto2002}.
Crystallization events in supercooled ST2 water are particularly important to study,
as it has been proposed that LDL is unstable against crystallization
\cite{Limmer2011}.

Following the discussion in Sec.~\ref{SEC:structural_parameters}, we
%Using the parameter $d_3^{(1)}(i,j)$ as discussed  we
%are able to measure the size of tiny crystals in the liquid.
%We 
define a crystal as a cluster of molecules which has three out of four
bonds with $d_3^{(1)} < -0.87$ and belong to 
the first coordination shell of each other.
In this section we shall study the growth and melting of these crystal nuclei, and estimate
the critical nucleus size needed to overcome the free energy barrier.
The existence of this barrier allows us to conclude that LDL is in
fact a {\it bona fide} metastable state with respect to the crystal.

\begin{figure}[htb]
    \centering
	\includegraphics[angle=0,width=0.45\textwidth]{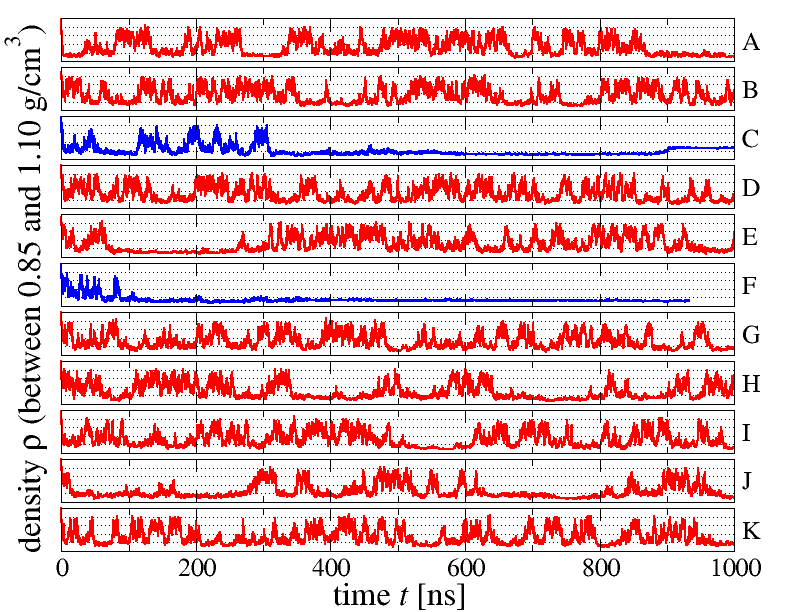}
    \caption{\label{FIG:rho_vs_time_with_crystals}
Density vs. time near the phase transition line at $P=205$~MPa and $T=246$~K for several different configurations of $N=343$ molecules.
This state point lies near the phase transition, and therefore phase flipping is seen to occur.
Runs C and F (partially) crystallize and, at that moment, cease to phase flip and remain stable at a low density.
    }
\end{figure}

\begin{figure}[htb]
    \centering
	\includegraphics[angle=0,width=0.45\textwidth]{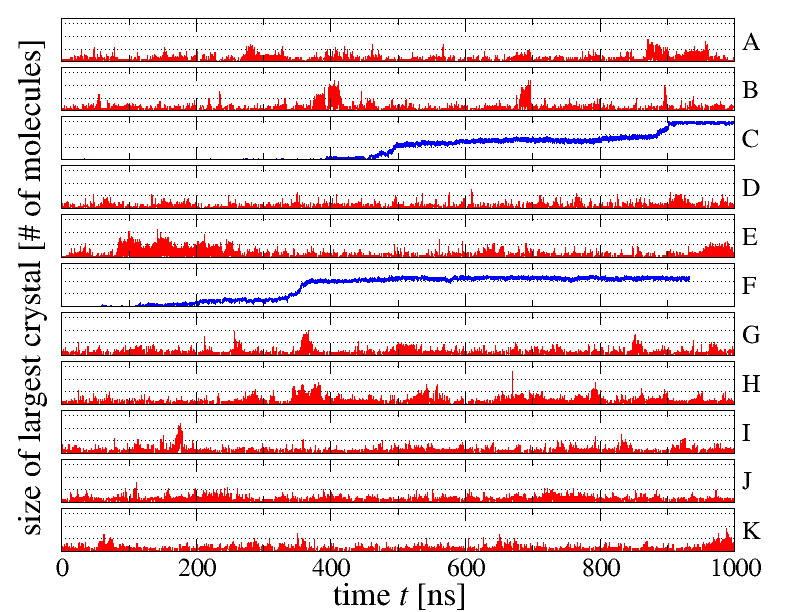}
    \caption{\label{FIG:crystalsize_vs_time}
Evolution of crystal size with time for the same configurations as in Fig.~\ref{FIG:rho_vs_time_with_crystals}.
The $y$-axis goes from 0 to 34, except for configurations C and F which go up to 343.
The system spontaneously crystallizes in both C and F, while the largest crystals in the remaining configurations never
reach a size larger than 30 molecules.
    }
\end{figure}

\begin{figure}[htb]
    \centering
    \includegraphics[angle=0,width=0.4\textwidth]{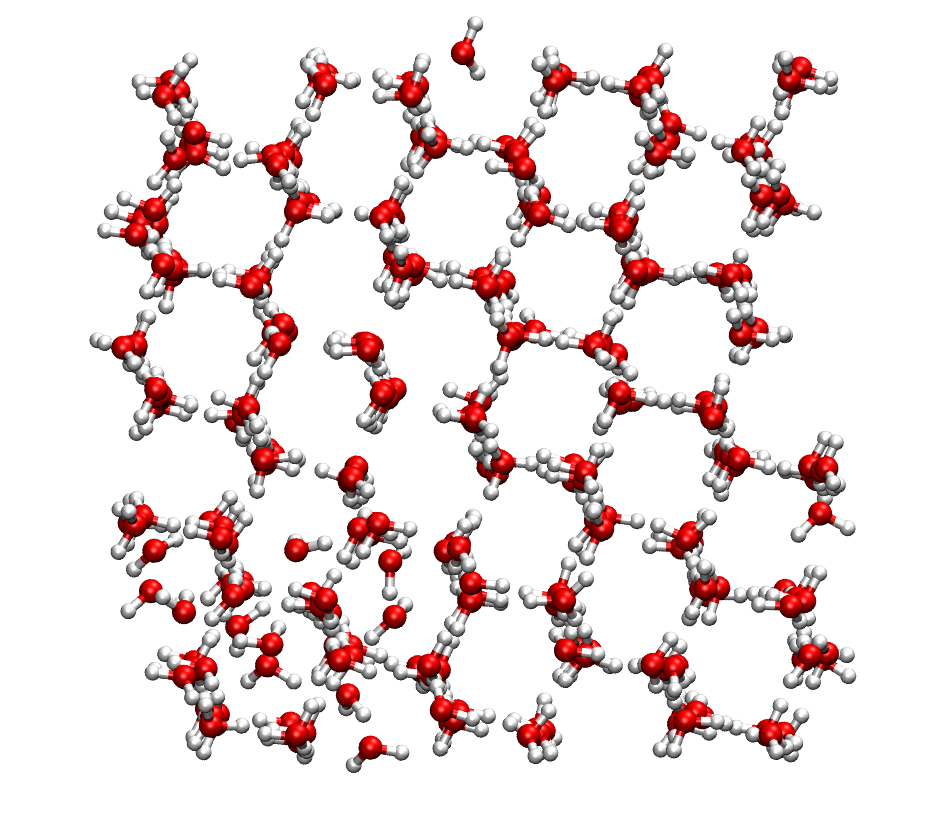}
    \caption{\label{FIG:crystal_snapshot}
A snapshot (at $t=1000$~ns) of the cubic diamond crystal produced by run C
of Figs.~\ref{FIG:rho_vs_time_with_crystals} and \ref{FIG:crystalsize_vs_time}.
Shown here are all $N=343$ molecules, with a small part still in the liquid state (bottom-left corner),
and a crystal defect in the center.
Note that the defect only affects the position of the hydrogen atoms, and not that of the oxygen.
    }
\end{figure}

In Fig.~\ref{FIG:rho_vs_time_with_crystals} we show the density evolution for 11 different configurations,
each with 343 molecules and at 205~MPa and 246~K.
Each of these runs started at a different initial density (between 0.85 and 0.95~g/cm$^3$) and was subsequently
equilibrated to the final temperature and pressure using the procedure described in Sec.~\ref{SEC:simulation_details}.
Because this state point lies close to the LLPT, we see phase flipping in all of them.
However, the two configurations C and F display a sudden jump to a stable low density plateau.
This is a hallmark of crystalization.
We confirm this by calculating the size of the largest crystal as a function of time (Fig.~\ref{FIG:crystalsize_vs_time}).
During most runs the largest crystal continuously grows and shrinks, but never reaches a size larger than 30 molecules.
On the other hand, configurations C and F show a jump in crystal size exactly matching the jump in density.
Run F ends up partially crystalized, while for C we find that over
90\% of the box is crystallized in a diamond structure with a density of about 0.92~g/cm$^3$
(Fig.~\ref{FIG:crystal_snapshot}).

\begin{figure}[htb]
    \centering
	\includegraphics[angle=0,width=0.45\textwidth]{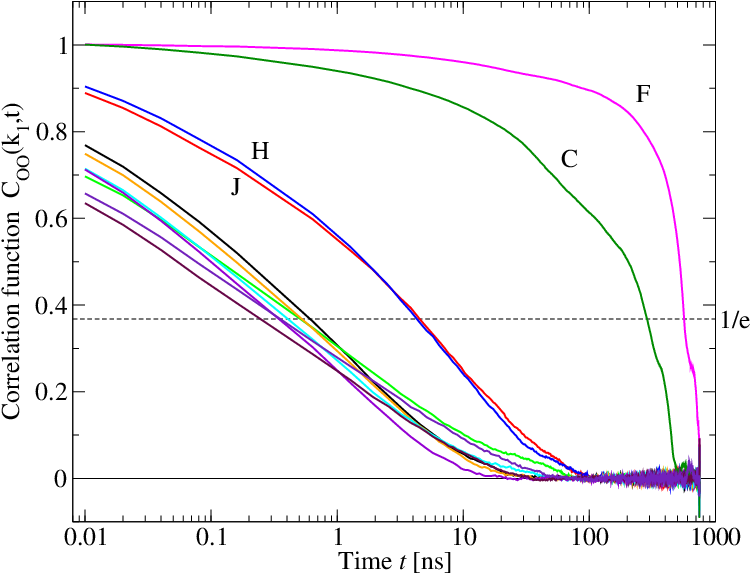}
    \caption{\label{FIG:tau_increase_with_crystals}
The correlation time increases dramatically if crystals of a size comparable
to the system size appear (i.e. runs C and F of Figs.~\ref{FIG:rho_vs_time_with_crystals} and \ref{FIG:crystalsize_vs_time}).
The correlation time of two other runs (H and J) are slightly larger than average because these runs
spend more time in the LDL phase (see Fig.~\ref{FIG:rho_vs_time_with_crystals}).
    }
\end{figure}

The correlation time increases dramatically if crystals appear with a size comparable
to the system size, as is evident from Fig.~\ref{FIG:tau_increase_with_crystals}.
The correlation functions of C and F decay very slowly, leading to correlation times of 200--400~ns,
while the other configurations have a correlation time of less than 4~ns.

For spontaneous crystallization to occur, a sufficiently large crystal nucleus needs to form within the liquid.
According to classical nucleation theory, this nucleus needs to reach a minimum size to prevent it from melting.
We observed in many simulations that a small nucleus grows and melts, and a few runs in which the nucleus grows further or remains stable.
Therefore, we can make an estimate of the critical nucleus size.

\begin{figure}[htb]
    \centering
	\includegraphics[angle=0,width=0.45\textwidth]{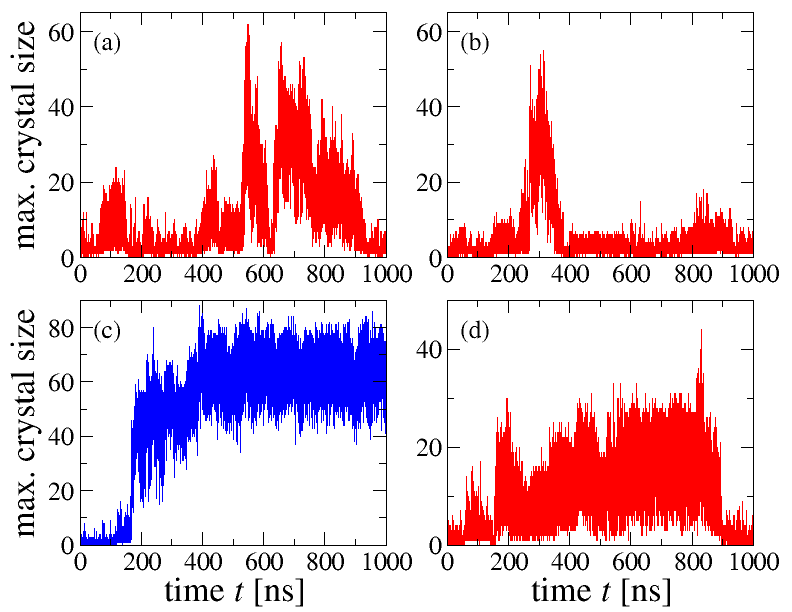}
    \caption{\label{FIG:crystal_growth_and_melting}
	Growth and melting of crystal nuclei.
(a) The largest nucleus that melted reached a size of 62 molecules during a simulation of 512 molecules at 210~MPa and 244~K.
(b) The second-largest nucleus was 55 molecules during a simulation of 343 molecules at 210~MPa and 243~K.
(c) A few runs lead to irreversible crystallization ($N=216$ at 195~MPa and 245~K).
(d) Some crystal nuclei survive for hundreds of nanoseconds ($N=343$
at 195~MPa and 246~K) before disappearing.
    }
\end{figure}

The two largest crystals that formed and subsequently melted, both
reached a size of about 50--60 molecules (Fig.~\ref{FIG:crystal_growth_and_melting}a and \ref{FIG:crystal_growth_and_melting}b).
The smallest crystal that formed and remained stable, had a size of about 50--80 molecules (Fig.~\ref{FIG:crystal_growth_and_melting}c).
We therefore conclude that the critical nucleus size is approximately $70 \pm 10$ molecules.
A similar value of $\simeq 85$ molecules was found by Reinhardt and Doye \cite{Reinhardt2012}
for ice nucleation in the monatomic water model \cite{Molinero2009}.

For a more accurate estimate it is necessary to run longer simulations,
as the crystal nuclei can survive for hundreds of nanoseconds
(e.g., Fig.~\ref{FIG:crystal_growth_and_melting}d in which a small crystal lasts for 700~ns).

%%%%%%%%%%%%%%%%%%%%%%%%%%%%%%%%%%%%%%%%%%%%%%%%%%%%%%%%%%%%%%%%%%%%%%%%%%%%%%%

\section{Locating the critical point
\label{SEC:locating_LLCP}}

\begin{figure}[htb]
    \centering
    \includegraphics[angle=0,width=0.45\textwidth]{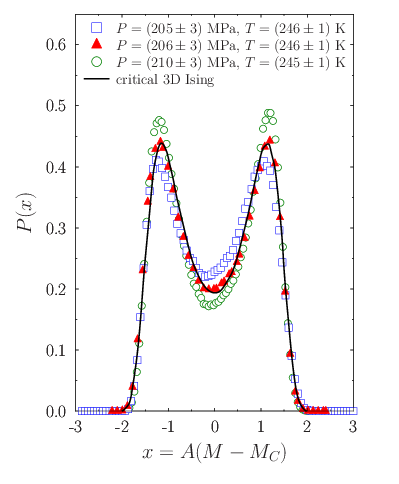}
    \caption{\label{FIG:order_parameter_fitting}
Fitting the order parameter distribution function (obtained from the simulation data) to that of the 3D Ising model at the critical point
(black curve, from Ref.~\cite{Hilfer1995}).
The shape of the order parameter distribution function depends on temperature and pressure.  Here (for $N=343$) we find an excellent fit
for $P_C=206$~MPa and $T_C=246$~K, and we therefore confirm that the LLCP indeed belongs to the same universality class as the 3D Ising model.
Based on our fit, we locate the LLCP to be at $P_C=206 \pm 3$~MPa and $T_C=246 \pm 1$~K.
    }
\end{figure}

In Sec.~\ref{SEC:Sk} we used the intermediate scattering function $S_{OO}(k)$ to estimate the position of the liquid-liquid critical point,
and found it to lie near 200--210~MPa and 244--247~K.
It is commonly believed that the LLCP falls in the same universality class as the three-dimensional Ising model \cite{Liu2009}.
At the critical point the order parameter distribution function (OPDF) of a system has the same bimodal shape as all other systems
that belong to the same universality class.
Therefore we can locate the LLCP accurately by fitting our data to the OPDF of the 3D Ising model (Fig.~\ref{FIG:order_parameter_fitting}).

In the 3D Ising model the order parameter $M$ is simply the spontaneous magnetization, but for liquids the order parameter
turns out to be a linear combination of two independent quantities such as the density and the potential energy \cite{Wilding1997,Bertrand2011}.
We therefore define $M \equiv \rho + s E$, with $s$ a constant known as the field mixing parameter.
The value of $s$ depends only on the model and should therefore be independent of the number of molecules.
Our fits indeed confirm this; we find $s=0.0362$~(g/cm$^3$)/(kJ/mol) for all values of $N$.

Only the shape of the OPDF is dictated by the theory, which means we are free to move and stretch our OPDF to acquire an accurate fit.
So, instead of fitting the order parameter $M$, we actually fit
$x\equiv A(M-M_C)$ to the 3D Ising model.
The critical order parameter $M_C$ is chosen such that the mean value of $x$ is zero, and the amplitude $A$ has been chosen such
that the variance equals unity.

\begin{figure}[htb]
    \centering
    \includegraphics[angle=0,width=0.45\textwidth]{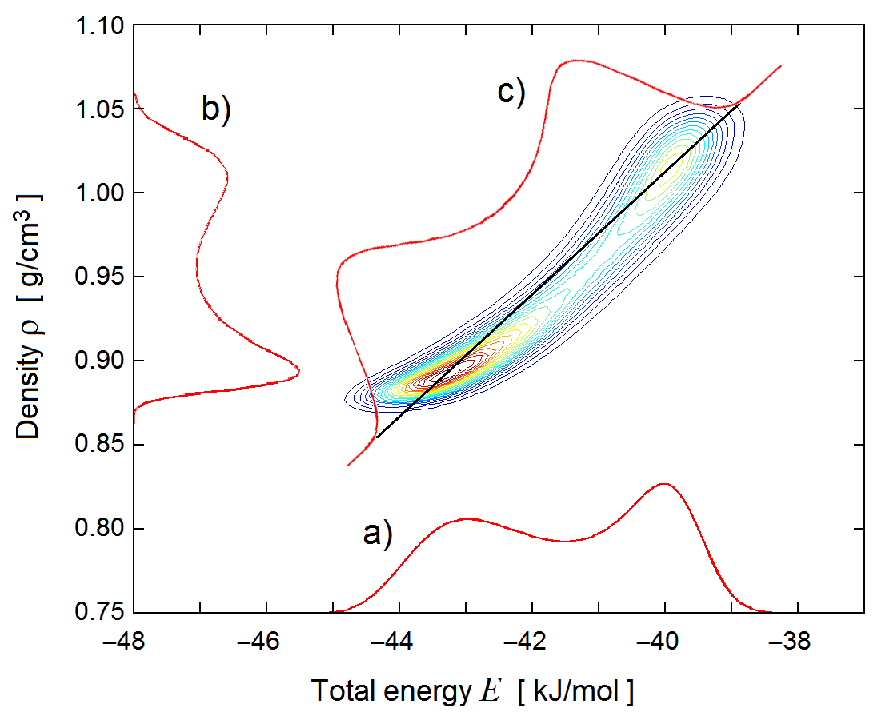}
    \caption{\label{FIG:2D_histogram_order_parameter}
2D histogram of the density and the total energy for a system at 247.5~K and 200~MPa (on the Widom line), obtained via histogram reweighting.
The histogram of the energy (curve a) seems to indicate that the system is mostly in the LDL state,
while the histogram of the density (curve b) indicates the HDL state is more predominant.
For liquids the order parameter $M\equiv \rho + s E$ is actually a linear combination of the density $\rho$ and the energy $E$ (curve c),
with $s=0.0362$~(g/cm$^3$)/(kJ/mol).
    }
\end{figure}

To calculate the OPDF for a particular pressure and temperature, we create a two-dimensional histogram of the density and energy.
An example of such a 2D histogram is shown in Fig.~\ref{FIG:2D_histogram_order_parameter} for $P=200$~MPa, $T=247.5$~K, and $N=343$.
Near the critical point the histogram displays two peaks, one for LDL and one for HDL.
If we integrate the 2D histogram along the direction corresponding to
the value of $s$, we obtain the histogram for $M$.

To fit our OPDF to that of the 3D Ising model, we need to calculate our OPDF at different pressures and temperatures,
until we find the $(P_C,T_C)$ that gives us the best fit.  The state point $(P_C,T_C)$ is then our best estimate of the location of the LLCP.
We only have simulation data for a finite number of state points, therefore some kind of interpolation is necessary.
The method of choice here is histogram reweighting \cite{Ferrenberg1989};
we use the algorithm as described by Panagiotopoulos in Ref.~\cite{Panagiotopoulos2000}.

\begin{figure}[htb]
    \centering
    \includegraphics[angle=0,width=0.45\textwidth]{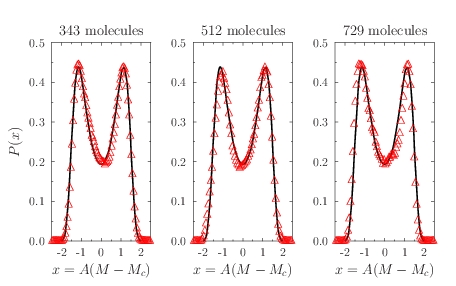}
    \caption{\label{FIG:order_parameter_vs_N}
Best OPDF fits for $N=343$, 512, and 729 molecules.  In all cases we obtain 
$s=0.0362$~(g/cm$^3$)/(kJ/mol) for the field mixing parameter.
We find the critical point to be located at $T_C=246 \pm 1$~K, $P_C=206 \pm 3$~MPa for $N=343$,
and at $T_C=246 \pm 1$~K, $P_C=208 \pm 3$~MPa for $N=529$ and 729.
    }
\end{figure}

\begin{table}[htb]
    \centering
\begin{tabular}{@{}cccccc @{}}
\hline
  & $190 \rm{MPa}$ & $195\rm{MPa}$ & $200\rm{MPa}$ & $205\rm{MPa}$& $210\rm{MPa}$\\
\hline
$242\rm{K}$& -  &  -   &   -   &  -  & 1\\
$243\rm{K}$& 1 &  1  &  1   & 1  & 1\\
$244\rm{K}$& 1 &  1  &  1   & 11& 1\\
$245\rm{K}$& 1 &  1  &  1   & 1  & 1\\
$246\rm{K}$& 1 & 11 & 10 & 9  & 1\\
$247\rm{K}$& 1 &  1  &  1   & 1  & 1\\
$248\rm{K}$& 1 &  1  &  2   & 1  & 1\\
$249\rm{K}$& 1 & 11 &  1   & 1  & 1\\
$250\rm{K}$& 1 &  1  &  1   & 1  & 1\\
$251\rm{K}$& 1 &  1  &  1   & 1  & 1\\
$252\rm{K}$& 1 & 11 &  1   & 1  & 1\\
\hline
\end{tabular}
	\caption{\label{TAB:hist_reweight_N343}
Number of simulations used for the histogram reweighting in order
to obtain the order parameter distribution function for $N=343$ molecules.
	}
\end{table}

\begin{table}[htb]
    \centering
\begin{tabular}{@{}ccccccccc@{}}
\hline
   &  & $N=512$ & & & $N=729$ & \\
   & 190MPa & 200MPa & 210MPa & 190MPa & 200MPa & 210MPa \\
\hline
242K &  -  &  -  &  4  &  -  &  -  &  -  \\
243K &  -  &  1  &  6  &  -  &  -  &  1  \\
244K &  -  &  1  &  5  &  -  &  -  &  1  \\
245K &  1  &  3  &  6  &  -  &  1  &  1  \\
246K &  1  &  2  &  6  &  1  &  1  &  1  \\
247K &  1  &  1  &  4  &  1  &  1  &  1  \\
248K &  1  &  2  &  4  &  1  &  1  &  1  \\
249K &  1  &  1  &  -  &  1  &  1  &  1  \\
250K &  1  &  -  &  -  &  1  &  1  &  -  \\
251K &  -  &  -  &  -  &  1  &  -  &  -  \\
\hline
\end{tabular}
	\caption{\label{TAB:hist_reweight_N512_N729}
Simulations used for the histogram reweighting
in order to obtain the order parameter distribution function for $N=512$ and
$N=729$ molecules.
	}
\end{table}

\begin{figure}[htb]
    \centering
	\includegraphics[angle=0,width=0.5\textwidth]{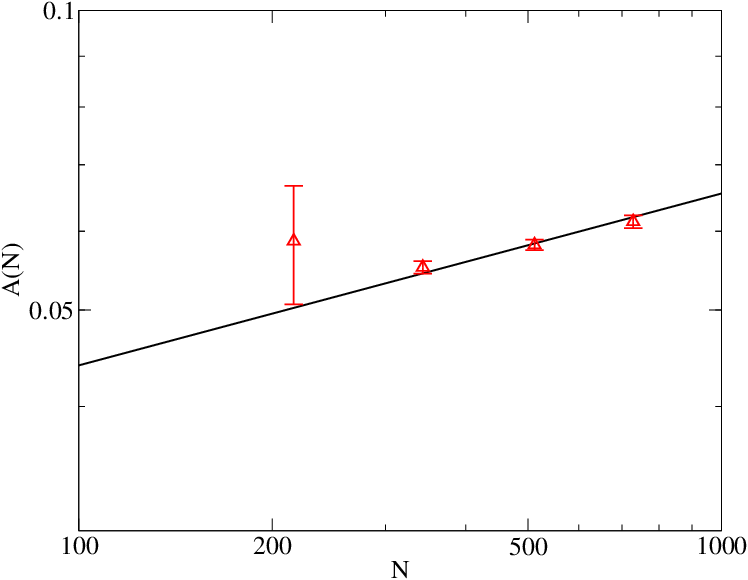}
    \caption{\label{FIG:order_parameter_fit_for_A}
Log-log plot of the amplitude $A$ vs. system size $N$.  From the slope of this line
we determine that $A \sim N^{0.16} \propto L^{0.48}$, in fare agreement with the
value of 0.52 predicted by the 3D Ising model \cite{Wilding1997}. For
the smaller size $N=216$ we observe large finite-size deviation with
respect to the thermodynamic limit behavior.
    }
\end{figure}

The results of fitting our data to the 3D Ising model are shown in Fig.~\ref{FIG:order_parameter_vs_N}.
Tables \ref{TAB:hist_reweight_N343} and \ref{TAB:hist_reweight_N512_N729}
indicate which data was used by the histogram reweighting method to obtain these fits.
For $N=343$, 512, and 729, we are able to fit our data very accurately to the OPDF of the 3D Ising model, and find
the critical point to be located at $T_C=246 \pm 1$~K, $P_C=206 \pm 3$~MPa for $N=343$,
and at $T_C=246 \pm 1$~K, $P_C=208 \pm 3$~MPa for $N=529$ and 729.
Theory predicts that the location of the critical point depends on $N$, and these findings agree with that prediction.
In particular, the 3D Ising model predicts that the amplitude $A$ should scale with box size $L$ as $A \sim L^{\beta/\nu} \propto N^{\beta/3\nu}$
with $\beta/\nu = 0.52$ \cite{Wilding1997, Odor2004}, in agreement with the slope of $A(N)$ in Fig.~\ref{FIG:order_parameter_fit_for_A}.
This figure also indicates that $N=216$ cannot provide an accurate estimate of the location of the LLCP.

To establish that the LLPT does not vanish in the thermodynamic limit $N \rightarrow \infty$,
we consider the finite size scaling of the Challa-Landau-Binder
parameter \cite{Challa1986, Franzese1998, Franzese2000a,Franzese2000b,Strekalova2011, Strekalova2012a}.
Near the critical point the density distribution function $\mathcal{D}(\rho)$ has a bimodal shape that can be approximated
by the superposition of two Gaussians (e.g., Fig.~\ref{FIG:2D_histogram_order_parameter}).
The Challa-Landau-Binder parameter $\Pi$ is a measure of the bimodality of $\mathcal{D}(\rho)$ and is defined as
\begin{align}
  \Pi \equiv 1 - \frac{\langle \rho^4 \rangle}{3 {\langle \rho^2 \rangle}^2}
\end{align}
When there is only one phase, $\mathcal{D}(\rho)$ is unimodal and $\Pi=2/3$.  But in a two-phase region,
with two phases that have different densities,
the shape of $\mathcal{D}(\rho)$ is bimodal (Fig.~\ref{FIG:2D_histogram_order_parameter}) and $\Pi < 2/3$.
For a finite system $\mathcal{D}(\rho)$ is always bimodal at both the Widom line and the LLPT, but in the thermodynamic limit
there exists only one phase at the Widom line, while there remain two
at the phase transition line. 
Therefore, $\Pi \rightarrow 2/3$ at the Widom line, while $\Pi < 2/3$ at the LLPT even in the limit $N \rightarrow \infty$.
Hence, the finite-size scaling of $\Pi$ allows us to distinguish whether an isobar crosses the LLPT or the Widom line, and is
yet another method of estimating the location of the critical point.

\begin{figure}[htb]
    \centering
    \includegraphics[angle=0,width=0.45\textwidth]{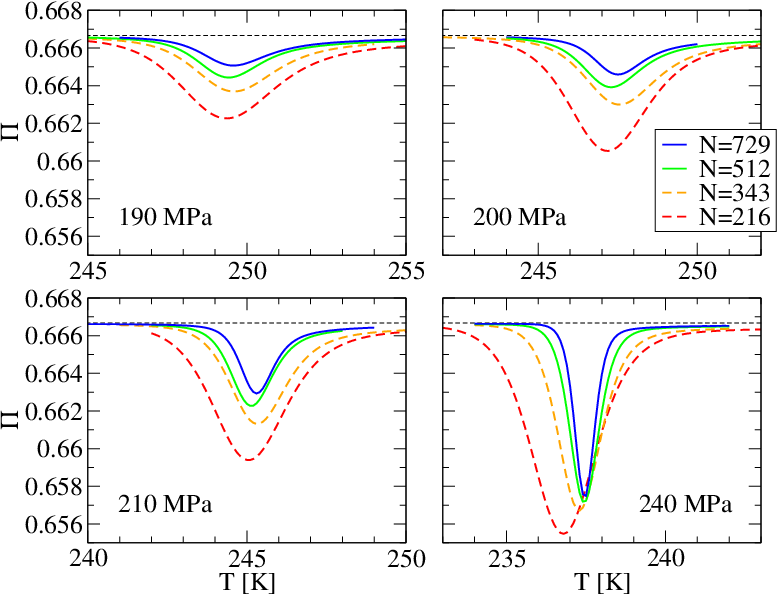}
    \caption{\label{FIG:CLB_parameter}
The Challa-Landau-Binder parameter $\Pi$ as a function of temperature and system size $N$, for four different pressures.
For finite system sizes $\Pi$ shows a minimum at the LLPT and the Widom line, while $\Pi \approx 2/3$ (thin dashed line)
at temperatures where $\mathcal{D}(\rho)$ is given by a single Gaussian.
The finite-size scaling of the minimum of $\Pi$, indicates that the critical point exists in the thermodynamic limit
(Fig.~\ref{FIG:minimum_of_CLB_parameter_vs_N}).
    }
\end{figure}

\begin{figure}[htb]
    \centering
    \includegraphics[angle=0,width=0.45\textwidth]{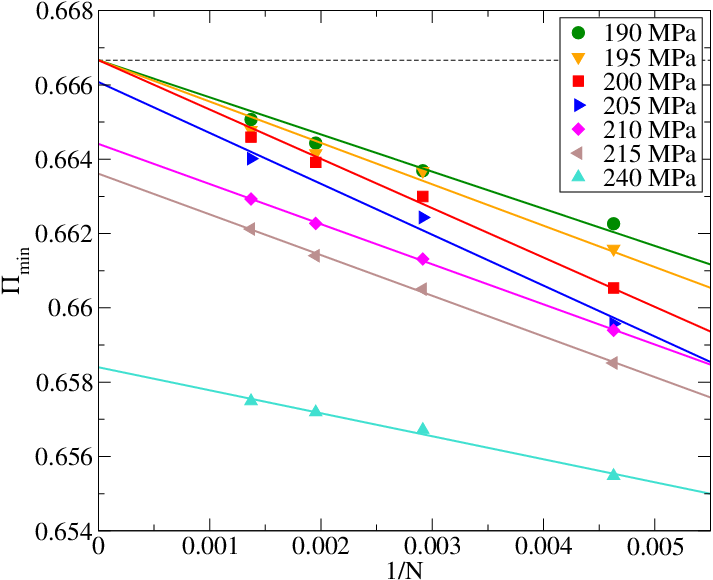}
    \caption{\label{FIG:minimum_of_CLB_parameter_vs_N}
Minima of the Challa-Landau-Binder parameter $\Pi$ as a function of system size $N$ for different pressures.
The minimum $\Pi_{\mathrm{ min}}$ occurs at the pressures and temperatures of the LLPT and the Widom line,
and is always less than $2/3$ for a finite system because of the bimodality of the density histogram.
As $N \rightarrow \infty$ the bimodality disappears in the one-phase region but remains at the LLPT,
and therefore $\Pi_{\mathrm{ min}} \rightarrow 2/3$ at the Widom line while $\Pi_{\mathrm{ min}} < 2/3$ on the LLPT, even in the thermodynamic limit.
We conclude that the critical point survives in the thermodynamic limit, and that it is located between $P=200$ and 210~MPa
(in agreement with previous results of this paper).
    }
\end{figure}

We study  
$\Pi$ versus temperature $T$ and system size $N$ for different
pressures, finding minima 
$\Pi_{\rm min}$ at specific temperature for each pressure (Fig.~\ref{FIG:CLB_parameter}).
The finite-size dependence of $\Pi_{\mathrm{min}}(P)$ reveals if
$P<P_C$ or $P>P_C$ (Fig.~\ref{FIG:minimum_of_CLB_parameter_vs_N}).

For $P < P_C$ the mimimum $\Pi_{\mathrm{min}}$ approaches $2/3$ linearly with $1/N$,
while for $P\leq P_C$ it 
approaches the limit \cite{Challa1986}
\begin{align}
  \Pi_{\mathrm{min}} \rightarrow  \frac{2}{3} - \frac{1}{3} \frac{
			( \rho_{\mathrm{H}}^2 - \rho_{\mathrm{L}}^2 )^2
		}{
			( \rho_{\mathrm{H}}^2 + \rho_{\mathrm{L}}^2 )^2.
		}
\end{align}
This limiting value is also approached linearly with $1/N$.
Here $\rho_{\mathrm{H}} \equiv \rho_{\mathrm{H}}(P)$ and $\rho_{\mathrm{L}} \equiv \rho_{\mathrm{L}}(P)$ are the
densities of the two phases LDL and HDL \cite{Franzese2000a}.
Above the critical pressure the limiting value of $\Pi_{\mathrm{min}}$ decreases as $P$ increases,
i.e. the two peaks of the bimodal $\mathcal{D}(\rho)$ move further apart.
This happens because $\rho_{\mathrm{H}} - \rho_{\mathrm{L}}$ increases at coexistence as $(P-P_C)^{\beta}$ where $\beta \approx 0.3$
is the critical exponent of the 3D Ising universality class \cite{Holten2012a,Holten2012b}.

From this analysis (Fig.~\ref{FIG:minimum_of_CLB_parameter_vs_N}) we conclude that
our results agree with theory and that the critical pressure $P_C \approx 190$--210~MPa, in agreement with the estimate of Sec.~\ref{SEC:locating_LLCP}.
Furthermore, as $\Pi$ remains less than $2/3$ for $P > P_C$ even in the limit $N \rightarrow \infty$, we conclude that
the LLPT does not vanish in the thermodynamic limit.

%%%%%%%%%%%%%%%%%%%%%%%%%%%%%%%%%%%%%%%%%%%%%%%%%%%%%%%%%%%%%%%%%%%%%%%%%%%%%%%

\section{Conclusions
\label{SEC:conclusions}}

We performed molecular dynamic simulations in the $NPT$ ensemble for ST2-RF water in the supercooled
region of the phase diagram for different system sizes with simulation times of up to 1000~ns.
Using several different techniques we confirmed the existence of two
liquid phases, LDL and HDL, separated by a liquid-liquid phase
transition line. 
Near the LLPT line the system continuously flips between the two phases.
Because of finite size effects this phenomenon also occurs near the Widom line, but by fitting the order parameter distribution function to that
of the 3D Ising model, we were able to accurately determine the location of the liquid-liquid critical point (at $T_C=246 \pm 1$~K, $P_C=208 \pm 3$~MPa).
Finite size scaling of the Challa-Landau-Binder parameter indicates that the critical point does not disappear in the thermodynamic limit.

Both phases have been confirmed to be 
{\it bona fide} 
metastable liquids that differ substantially in structural as well as dynamical properties.
It is found that the LDL phase is a more ``structured'' liquid,
and that it has a correlation time of almost four orders of magnitude
larger than that of HDL, with LDL correlation time 
of the order of 100--1000~ns. We show that $Q_6$ structural parameter
is not able to discriminate between HDL and LDL, but can discriminate
well between liquids and crystal. Finite size scaling of the $Q_6$
parameter confirms  
that LDL scales as a liquid and not as a crystal.

The different structures of LDL and
HDL are better discriminated by structural parameters like $q_3^{(1)}$
and $q_6^{(2)}$. These parameters show that LDL and
HDL differ mostly in the amount of diamond structure of
the first shell and the amount of hcp structure in the second shell.

For small box sizes ($N = 343$) there were a few simulation runs that resulted in spontaneous crystallization,
always within the LDL region of the phase diagram.
Further analysis revealed that during all simulations small crystals grow and melt within the liquid, a clear indication that LDL is metastable
with respect to the crystal.
From the few crystalization events that occurred, we were able to conclude that the critical nucleus size is approximately $70 \pm 10$ molecules.

%%%%%%%%%%%%%%%%%%%%%%%%%%%%%%%%%%%%%%%%%%%%%%%%%%%%%%%%%%%%%%%%%%%%%%%%%%%%%%%

\section{Acknowledgements}

We thank Y. Liu, A. Z. Panagiotopoulos, P. Debenedetti, F. Sciortino, I. Saika-Voivod and P. H. Poole
for sharing their results, obtained using approaches different from ours, but also
addressing the LLCP hypothesis.
GF thanks Spanish MEC grant FIS2012-31025 co-financed FEDER and EU FP7 grant NMP4-SL-2011-266737 for support.
SVB acknowledges the partial support of this research through the Dr. Bernard W. Gamson Computational Science Center at Yeshiva College
and through the Departament d'Universitats, Recerca i Societat de la Informaci\'{o} de la Generalitat de Catalunya.
HES thanks the NSF Chemistry Division for support (grants CHE 0911389 and CHE 0908218).
HJH thanks the European Research Council (ERC) Advanced Grant 319968-FlowCCS.

%%%%%%%%%%%%%%%%%%%%%%%%%%%%%%%%%%%%%%%%%%%%%%%%%%%%%%%%%%%%%%%%%%%%%%%%%%%%%%%

%\bibliographystyle{jcp}  % use style file "jcp.bst"
%\bibliographystyle{aip}  % use style file "aip.bst"

%\bibliography{references}

\begin{thebibliography}{100}

\bibitem{Angell1973}
{\sc C.~A. Angell}, {\sc J.~Shuppert}, and {\sc J.~C. Tucker},
\newblock {\em J. Phys. Chem.} {\bf 77}, 3092 (1973).

\bibitem{Speedy1976}
{\sc R.~J. Speedy} and {\sc C.~A. Angell},
\newblock {\em J. Chem. Phys.} {\bf 65}, 851 (1976).

\bibitem{Angell1982}
{\sc C.~A. Angell}, {\sc W.~J. Sichina}, and {\sc M.~Oguni},
\newblock {\em J. Phys. Chem.} {\bf 86}, 998 (1982).

\bibitem{Speedy1982}
{\sc R.~J. Speedy},
\newblock {\em J. Phys. Chem.} {\bf 86}, 982 (1982).

\bibitem{Burton1935}
{\sc E.~F. Burton} and {\sc W.~F. Oliver},
\newblock {\em Proc. R. Soc. A} {\bf 153}, 166 (1935).

\bibitem{Bruggeller1980}
{\sc P.~Br\"{u}ggeller} and {\sc E.~Mayer},
\newblock {\em Nature} {\bf 288}, 569 (1980).

\bibitem{Mishima1984}
{\sc O.~Mishima}, {\sc L.~D. Calvert}, and {\sc E.~Whalley},
\newblock {\em Nature} {\bf 310}, 393 (1984).

\bibitem{Loerting2006}
{\sc T.~Loerting} and {\sc N.~Giovambattista},
\newblock {\em J. Phys.: Condens. Matter} {\bf 18}, R919 (2006).

\bibitem{Mishima1985}
{\sc O.~Mishima}, {\sc L.~D. Calvert}, and {\sc E.~Whalley},
\newblock {\em Nature} {\bf 314}, 76 (1985).

\bibitem{Mishima1994}
{\sc O.~Mishima},
\newblock {\em J. Chem. Phys.} {\bf 100}, 5910 (1994).

\bibitem{Mishima1998a}
{\sc O.~Mishima} and {\sc H.~E. Stanley},
\newblock {\em Nature} {\bf 392}, 164 (1998).

\bibitem{Mishima1998b}
{\sc O.~Mishima} and {\sc H.~E. Stanley},
\newblock {\em Nature} {\bf 396}, 329 (1998).

\bibitem{Nilsson2012pc}
{\sc A.~Nilsson},
\newblock -,
\newblock private communication, 2012.

\bibitem{Poole1992}
{\sc P.~H. Poole}, {\sc F.~Sciortino}, {\sc U.~Essmann}, and {\sc H.~E.
  Stanley},
\newblock {\em Nature} {\bf 360}, 324 (1992).

\bibitem{Stillinger1974}
{\sc F.~H. Stillinger} and {\sc A.~Rahman},
\newblock {\em J. Chem. Phys.} {\bf 60}, 1545 (1974).

\bibitem{Tokushima2008}
{\sc T.~Tokushima}, {\sc Y.~Harada}, {\sc O.~Takahashi}, {\sc Y.~Senba}, {\sc
  H.~Ohashi}, {\sc L.~G.~M. Pettersson}, {\sc A.~Nilsson}, and {\sc S.~Shin},
\newblock {\em Chem. Phys. Lett.} {\bf 460}, 387 (2008).

\bibitem{Huang2009}
{\sc C.~Huang}, {\sc K.~T. Wikfeldt}, {\sc T.~Tokushima}, {\sc D.~Nordlund},
  {\sc Y.~Harada}, {\sc U.~Bergmann}, {\sc M.~Niebuhr}, {\sc T.~M. Weiss}, {\sc
  Y.~Horikawa}, {\sc M.~Leetmaa}, {\sc M.~P. Ljungberg}, {\sc O.~Takahashi},
  {\sc A.~Lenz}, {\sc L.~Ojam\"{a}e}, {\sc A.~P. Lyubartsev}, {\sc S.~Shin},
  {\sc L.~G.~M. Pettersson}, and {\sc A.~Nilsson},
\newblock {\em Proc. Natl. Acad. Sci. U.S.A.} {\bf 106}, 15214 (2009).

\bibitem{Nilsson2011}
{\sc A.~Nilsson} and {\sc L.~G.~M. Pettersson},
\newblock {\em Chem. Phys.} {\bf 389}, 1 (2011).

\bibitem{Wikfeldt2011}
{\sc K.~T. Wikfeldt}, {\sc A.~Nilsson}, and {\sc L.~G.~M. Pettersson},
\newblock {\em Phys. Chem. Chem. Phys.} {\bf 13}, 19918 (2011).

\bibitem{Huang2010}
{\sc C.~Huang}, {\sc T.~M. Weiss}, {\sc D.~Nordlund}, {\sc K.~T. Wikfeldt},
  {\sc L.~G.~M. Pettersson}, and {\sc A.~Nilsson},
\newblock {\em J. Chem. Phys.} {\bf 133}, 134504 (2010).

\bibitem{Zhang2011}
{\sc Y.~Zhang}, {\sc A.~Faraone}, {\sc W.~A. Kamitakahara}, {\sc K.-H. Liu},
  {\sc C.-Y. Mou}, {\sc J.~B. Le\~{a}o}, {\sc S.~Chang}, and {\sc S.-H. Chen},
\newblock {\em Proc. Natl. Acad. Sci. U.S.A.} {\bf 108}, 12206 (2011).

\bibitem{Mazza2011}
{\sc M.~G. Mazza}, {\sc K.~Stokely}, {\sc S.~E. Pagnotta}, {\sc F.~Bruni}, {\sc
  H.~E. Stanley}, and {\sc G.~Franzese},
\newblock {\em Proc. Natl. Acad. Sci. U.S.A.} {\bf 108}, 19873 (2011).

\bibitem{Franzese2011}
{\sc G.~Franzese}, {\sc V.~Bianco}, and {\sc S.~Iskrov},
\newblock {\em Food Biophys.} {\bf 6}, 186 (2011).

\bibitem{Bianco2012a}
{\sc V.~Bianco}, {\sc S.~Iskrov}, and {\sc G.~Franzese},
\newblock {\em J. Biol. Phys.} {\bf 38}, 27 (2012).

\bibitem{Kumar2011}
{\sc P.~Kumar} and {\sc H.~E. Stanley},
\newblock {\em J. Phys. Chem. B} {\bf 115}, 14269 (2011).

\bibitem{Sciortino1990}
{\sc F.~Sciortino}, {\sc P.~H. Poole}, {\sc H.~E. Stanley}, and {\sc
  S.~Havlin},
\newblock {\em Phys. Rev. Lett.} {\bf 64}, 1686 (1990).

\bibitem{Starr1999a}
{\sc F.~W. Starr}, {\sc J.~K. Nielsen}, and {\sc H.~E. Stanley},
\newblock {\em Phys. Rev. Lett.} {\bf 82}, 2294 (1999).

\bibitem{Kumar2008a}
{\sc P.~Kumar}, {\sc G.~Franzese}, and {\sc H.~E. Stanley},
\newblock {\em Phys. Rev. Lett.} {\bf 100}, 105701 (2008).

\bibitem{Kumar2008b}
{\sc P.~Kumar}, {\sc G.~Franzese}, and {\sc H.~E. Stanley},
\newblock {\em J. Phys.: Condens. Matter} {\bf 20}, 244114 (2008).

\bibitem{Franzese2009}
{\sc G.~Franzese} and {\sc F.~de~los Santos},
\newblock {\em J. Phys.: Condens. Matter} {\bf 21}, 504107 (2009).

\bibitem{delosSantos2011}
{\sc F.~de~los Santos} and {\sc G.~Franzese},
\newblock {\em J. Phys. Chem. B} {\bf 115}, 14311 (2011).

\bibitem{delosSantos2012}
{\sc F.~de~los Santos} and {\sc G.~Franzese},
\newblock {\em Phys. Rev. E} {\bf 85}, 010602 (2012).

\bibitem{Mazza2012}
{\sc M.~G. Mazza}, {\sc K.~Stokely}, {\sc H.~E. Stanley}, and {\sc
  G.~Franzese},
\newblock {\em J. Chem. Phys.} {\bf 137}, 204502 (2012).

\bibitem{Franzese2008}
{\sc G.~Franzese}, {\sc K.~Stokely}, {\sc X.~Q. Chu}, {\sc P.~Kumar}, {\sc
  M.~G. Mazza}, {\sc S.~H. Chen}, and {\sc H.~E. Stanley},
\newblock {\em J. Phys.: Condens. Matter} {\bf 20}, 494210 (2008).

\bibitem{Stanley2009}
{\sc H.~E. Stanley}, {\sc P.~Kumar}, {\sc S.~Han}, {\sc M.~G. Mazza}, {\sc
  K.~Stokely}, {\sc S.~V. Buldyrev}, {\sc G.~Franzese}, {\sc F.~Mallamace}, and
  {\sc L.~Xu},
\newblock {\em J. Phys.: Condens. Matter} {\bf 21}, 504105 (2009).

\bibitem{Stanley2010}
{\sc H.~E. Stanley}, {\sc S.~V. Buldyrev}, {\sc G.~Franzese}, {\sc P.~Kumar},
  {\sc F.~Mallamace}, {\sc M.~G. Mazza}, {\sc K.~Stokely}, and {\sc L.~Xu},
\newblock {\em J. Phys.: Condens. Matter} {\bf 22}, 284101 (2010).

\bibitem{Stanley2011}
{\sc H.~E. Stanley}, {\sc S.~V. Buldyrev}, {\sc P.~Kumar}, {\sc F.~Mallamace},
  {\sc M.~G. Mazza}, {\sc K.~Stokely}, {\sc L.~Xu}, and {\sc G.~Franzese},
\newblock {\em J. Non-Cryst. Solids} {\bf 357}, 629 (2011).

\bibitem{Harrington1997}
{\sc S.~Harrington}, {\sc P.~H. Poole}, {\sc F.~Sciortino}, and {\sc H.~E.
  Stanley},
\newblock {\em J. Chem. Phys.} {\bf 107}, 7443 (1997).

\bibitem{Franzese2001}
{\sc G.~Franzese}, {\sc G.~Malescio}, {\sc A.~Skibinsky}, {\sc S.~V. Buldyrev},
  and {\sc H.~E. Stanley},
\newblock {\em Nature} {\bf 409}, 692 (2001).

\bibitem{Franzese2003}
{\sc G.~Franzese}, {\sc M.~I. Marqu\'{e}s}, and {\sc H.~E. Stanley},
\newblock {\em Phys. Rev. E} {\bf 67}, 011103 (2003).

\bibitem{Franzese2007a}
{\sc G.~Franzese},
\newblock {\em J. Mol. Liq.} {\bf 136}, 267 (2007).

\bibitem{Hsu2008}
{\sc C.~W. Hsu}, {\sc J.~Largo}, {\sc F.~Sciortino}, and {\sc F.~W. Starr},
\newblock {\em Proc. Natl. Acad. Sci. U.S.A.} {\bf 105}, 13711 (2008).

\bibitem{Stanley2008}
{\sc H.~E. Stanley}, {\sc P.~Kumar}, {\sc G.~Franzese}, {\sc L.~Xu}, {\sc
  Z.~Yan}, {\sc M.~G. Mazza}, {\sc S.~V. Buldyrev}, {\sc S.-H. Chen}, and {\sc
  F.~Mallamace},
\newblock {\em Eur. Phys. J. Special Topics} {\bf 161}, 1 (2008).

\bibitem{Oliveira2008}
{\sc A.~B. de~Oliveira}, {\sc G.~Franzese}, {\sc P.~A. Netz}, and {\sc M.~C.
  Barbosa},
\newblock {\em The Journal of Chemical Physics} {\bf 128}, 064901 (2008).

\bibitem{Mazza2009}
{\sc M.~G. Mazza}, {\sc K.~Stokely}, {\sc E.~G. Strekalova}, {\sc H.~E.
  Stanley}, and {\sc G.~Franzese},
\newblock {\em Comp. Phys. Comm.} {\bf 180}, 497 (2009).

\bibitem{Franzese2010}
{\sc G.~Franzese}, {\sc A.~Hernando-Mart{\'\i}nez}, {\sc P.~Kumar}, {\sc M.~G.
  Mazza}, {\sc K.~Stokely}, {\sc E.~G. Strekalova}, {\sc F.~de~los Santos}, and
  {\sc H.~E. Stanley},
\newblock {\em J. Phys.: Condens. Matter} {\bf 22}, 284103 (2010).

\bibitem{Stokely2010}
{\sc K.~Stokely}, {\sc M.~G. Mazza}, {\sc H.~E. Stanley}, and {\sc
  G.~Franzese},
\newblock {\em Proc. Natl. Acad. Sci. U.S.A.} {\bf 107}, 1301 (2010).

\bibitem{Corradini2010a}
{\sc D.~Corradini}, {\sc M.~Rovere}, and {\sc P.~Gallo},
\newblock {\em J. Chem. Phys.} {\bf 132}, 134508 (2010).

\bibitem{Vilaseca2010}
{\sc P.~Vilaseca} and {\sc G.~Franzese},
\newblock {\em J. Chem. Phys.} {\bf 133}, 084507 (2010).

\bibitem{Vilaseca2011}
{\sc P.~Vilaseca} and {\sc G.~Franzese},
\newblock {\em J. Non-Cryst. Solids} {\bf 357}, 419 (2011).

\bibitem{Xu2011}
{\sc L.~Xu}, {\sc N.~Giovambattista}, {\sc S.~V. Buldyrev}, {\sc P.~G.
  Debenedetti}, and {\sc H.~E. Stanley},
\newblock {\em J. Chem. Phys.} {\bf 134}, 064507 (2011).

\bibitem{Gallo2012}
{\sc P.~Gallo} and {\sc F.~Sciortino},
\newblock {\em Phys. Rev. Lett.} {\bf 109}, 177801 (2012).

\bibitem{Strekalova2012b}
{\sc E.~G. Strekalova}, {\sc D.~Corradini}, {\sc M.~G. Mazza}, {\sc S.~V.
  Buldyrev}, {\sc P.~Gallo}, {\sc G.~Franzese}, and {\sc H.~E. Stanley},
\newblock {\em J. Biol. Phys.} {\bf 38}, 97 (2012).

\bibitem{Strekalova2012c}
{\sc E.~G. Strekalova}, {\sc J.~Luo}, {\sc H.~E. Stanley}, {\sc G.~Franzese},
  and {\sc S.~V. Buldyrev},
\newblock {\em Phys. Rev. Lett.} {\bf 109}, 105701 (2012).

\bibitem{Bianco2012b}
{\sc V.~Bianco} and {\sc G.~Franzese},
\newblock {\em arXiv:cond-mat.soft}  (2012).

\bibitem{Poole2005}
{\sc P.~H. Poole}, {\sc I.~Saika-Voivod}, and {\sc F.~Sciortino},
\newblock {\em J. Phys.: Condens. Matter} {\bf 17}, L431 (2005).

\bibitem{Liu2009}
{\sc Y.~Liu}, {\sc A.~Z. Panagiotopoulos}, and {\sc P.~G. Debenedetti},
\newblock {\em J. Chem. Phys.} {\bf 131}, 104508 (2009).

\bibitem{Liu2010}
{\sc Y.~Liu}, {\sc A.~Z. Panagiotopoulos}, and {\sc P.~G. Debenedetti},
\newblock {\em J. Chem. Phys.} {\bf 132}, 144107 (2010).

\bibitem{Yamada2002}
{\sc M.~Yamada}, {\sc S.~Mossa}, {\sc H.~E. Stanley}, and {\sc F.~Sciortino},
\newblock {\em Phys. Rev. Lett.} {\bf 88}, 195701 (2002).

\bibitem{Paschek2008}
{\sc D.~Paschek}, {\sc A.~R\"{u}ppert}, and {\sc A.~Geiger},
\newblock {\em ChemPhysChem} {\bf 9}, 2737 (2008).

\bibitem{Abascal2010}
{\sc J.~L.~F. Abascal} and {\sc C.~Vega},
\newblock {\em J. Chem. Phys.} {\bf 133}, 234502 (2010).

\bibitem{Abascal2011}
{\sc J.~L.~F. Abascal} and {\sc C.~Vega},
\newblock {\em J. Chem. Phys.} {\bf 134}, 186101 (2011).

\bibitem{Limmer2011}
{\sc D.~T. Limmer} and {\sc D.~Chandler},
\newblock {\em J. Chem. Phys.} {\bf 135}, 134503 (2011).

\bibitem{Poole2011}
{\sc P.~H. Poole}, {\sc S.~R. Becker}, {\sc F.~Sciortino}, and {\sc F.~W.
  Starr},
\newblock {\em J. Phys. Chem. B} {\bf 115}, 14176 (2011).

\bibitem{Kesselring2012}
{\sc T.~A. Kesselring}, {\sc G.~Franzese}, {\sc S.~V. Buldyrev}, {\sc H.~J.
  Herrmann}, and {\sc H.~E. Stanley},
\newblock {\em Sci. Rep.} {\bf 2}, 474 (2012).

\bibitem{Sciortino2011}
{\sc F.~Sciortino}, {\sc I.~Saika-Voivod}, and {\sc P.~H. Poole},
\newblock {\em Phys. Chem. Chem. Phys.} {\bf 13}, 19759 (2011).

\bibitem{Poole2013}
{\sc P.~H. Poole}, {\sc R.~K. Bowles}, {\sc I.~Saika-Voivod}, and {\sc
  F.~Sciortino},
\newblock {\em J. Chem. Phys.} {\bf 138}, 034505 (2013).

\bibitem{Liu2012}
{\sc Y.~Liu}, {\sc J.~C. Palmer}, {\sc A.~Z. Panagiotopoulos}, and {\sc P.~G.
  Debenedetti},
\newblock {\em J. Chem. Phys.} {\bf 137}, 214505 (2012).

\bibitem{Xu2005}
{\sc L.~Xu}, {\sc P.~Kumar}, {\sc S.~V. Buldyrev}, {\sc S.-H. Chen}, {\sc P.~H.
  Poole}, {\sc F.~Sciortino}, and {\sc H.~E. Stanley},
\newblock {\em Proc. Natl. Acad. Sci. U.S.A.} {\bf 102}, 16558 (2005).

\bibitem{Franzese2007b}
{\sc G.~Franzese} and {\sc H.~E. Stanley},
\newblock {\em J. Phys.: Condens. Matter} {\bf 19}, 205126 (2007).

\bibitem{Steinhauser1982}
{\sc O.~Steinhauser},
\newblock {\em Mol. Phys.} {\bf 45}, 335 (1982).

\bibitem{Horn2004}
{\sc H.~W. Horn}, {\sc W.~C. Swope}, {\sc J.~W. Pitera}, {\sc J.~D. Madura},
  {\sc T.~J. Dick}, {\sc G.~L. Hura}, and {\sc T.~Head-Gordon},
\newblock {\em J. Chem. Phys.} {\bf 120}, 9665 (2004).

\bibitem{Allen1987}
{\sc M.~P. Allen} and {\sc D.~J. Tildesley},
\newblock {\em Computer Simulation of Liquids},
\newblock Oxford Science Publications, 1987.

\bibitem{Ryckaert1977}
{\sc J.-P. Ryckaert}, {\sc G.~Ciccotti}, and {\sc H.~J.~C. Berendsen},
\newblock {\em J. Comp. Phys.} {\bf 23}, 327 (1977).

\bibitem{Nose1984}
{\sc S.~Nos\'{e}},
\newblock {\em J. Chem. Phys.} {\bf 81}, 511 (1984).

\bibitem{Nose1991}
{\sc S.~Nos\'{e}},
\newblock {\em Prog. Theor. Phys. Suppl.} {\bf 103}, 1 (1991).

\bibitem{Berendsen1984}
{\sc H.~J.~C. Berendsen}, {\sc J.~P.~M. Postma}, {\sc W.~F. van Gunsteren},
  {\sc A.~DiNola}, and {\sc J.~R. Haak},
\newblock {\em J. Chem. Phys.} {\bf 81}, 3684 (1984).

\bibitem{Franzese2002}
{\sc G.~Franzese}, {\sc G.~Malescio}, {\sc A.~Skibinsky}, {\sc S.~V. Buldyrev},
  and {\sc H.~E. Stanley},
\newblock {\em Phys. Rev. E} {\bf 66}, 051206 (2002).

\bibitem{Soper2000}
{\sc A.~K. Soper},
\newblock {\em Chem. Phys.} {\bf 258}, 121 (2000).

\bibitem{Starr1999b}
{\sc F.~W. Starr}, {\sc F.~Sciortino}, and {\sc H.~E. Stanley},
\newblock {\em Phys. Rev. E} {\bf 60}, 6757 (1999).

\bibitem{Kumar2006}
{\sc P.~Kumar}, {\sc G.~Franzese}, {\sc S.~V. Buldyrev}, and {\sc H.~E.
  Stanley},
\newblock {\em Phys. Rev. E} {\bf 73}, 041505 (2006).

\bibitem{Gallo1996}
{\sc P.~Gallo}, {\sc F.~Sciortino}, {\sc P.~Tartaglia}, and {\sc S.-H. Chen},
\newblock {\em Phys. Rev. Lett.} {\bf 76}, 2730 (1996).

\bibitem{Steinhardt1983}
{\sc P.~J. Steinhardt}, {\sc D.~R. Nelson}, and {\sc M.~Ronchetti},
\newblock {\em Phys. Rev. B} {\bf 28}, 784 (1983).

\bibitem{Ghiringhelli2008}
{\sc L.~M. Ghiringhelli}, {\sc C.~Valeriani}, {\sc J.~H. Los}, {\sc E.~J.
  Meijer}, {\sc A.~Fasolino}, and {\sc D.~Frenkel},
\newblock {\em Mol. Phys.} {\bf 106}, 2011 (2008).

\bibitem{Matsumoto2002}
{\sc M.~Matsumoto}, {\sc S.~Saito}, and {\sc I.~Ohmine},
\newblock {\em Nature} {\bf 416}, 409 (2002).

\bibitem{Reinhardt2012}
{\sc A.~Reinhardt} and {\sc J.~P.~K. Doye},
\newblock {\em J. Chem. Phys.} {\bf 136}, 054501 (2012).

\bibitem{Molinero2009}
{\sc V.~Molinero} and {\sc E.~B. Moore},
\newblock {\em J. Phys. Chem. B} {\bf 113}, 4008 (2009).

\bibitem{Hilfer1995}
{\sc R.~Hilfer} and {\sc N.~B. Wilding},
\newblock {\em J. Phys. A: Math. Gen.} {\bf 28}, L281 (1995).

\bibitem{Wilding1997}
{\sc N.~B. Wilding},
\newblock {\em J. Phys.: Condens. Matter} {\bf 9}, 585 (1997).

\bibitem{Bertrand2011}
{\sc C.~E. Bertrand} and {\sc M.~A. Anisimov},
\newblock {\em J. Phys. Chem. B} {\bf 115}, 14099 (2011).

\bibitem{Ferrenberg1989}
{\sc A.~M. Ferrenberg} and {\sc R.~H. Swendsen},
\newblock {\em Phys. Rev. Lett.} {\bf 63}, 1195 (1989).

\bibitem{Panagiotopoulos2000}
{\sc A.~Z. Panagiotopoulos},
\newblock {\em J. Phys.: Condens. Matter} {\bf 12}, 25 (2000).

\bibitem{Odor2004}
{\sc G.~\'{O}dor},
\newblock {\em Rev. Mod. Phys.} {\bf 76}, 663 (2004).

\bibitem{Challa1986}
{\sc M.~S.~S. Challa}, {\sc D.~P. Landau}, and {\sc K.~Binder},
\newblock {\em Phys. Rev. B} {\bf 34}, 1841 (1986).

\bibitem{Franzese1998}
{\sc G.~Franzese} and {\sc A.~Coniglio},
\newblock {\em Phys. Rev. E} {\bf 58}, 2753 (1998).

\bibitem{Franzese2000a}
{\sc G.~Franzese},
\newblock {\em Phys. Rev. E} {\bf 61}, 6383 (2000).

\bibitem{Franzese2000b}
{\sc G.~Franzese}, {\sc V.~Cataudella}, {\sc S.~E. Korshunov}, and {\sc
  R.~Fazio},
\newblock {\em Phys. Rev. B} {\bf 62}, R9287 (2000).

\bibitem{Strekalova2011}
{\sc E.~G. Strekalova}, {\sc M.~G. Mazza}, {\sc H.~E. Stanley}, and {\sc
  G.~Franzese},
\newblock {\em Phys. Rev. Lett.} {\bf 106}, 145701 (2011).

\bibitem{Strekalova2012a}
{\sc E.~G. Strekalova}, {\sc M.~G. Mazza}, {\sc H.~E. Stanley}, and {\sc
  G.~Franzese},
\newblock {\em J. Phys.: Condens. Matter} {\bf 24}, 064111 (2012).

\bibitem{Holten2012a}
{\sc V.~Holten}, {\sc C.~E. Bertrand}, {\sc M.~A. Anisimov}, and {\sc J.~V.
  Sengers},
\newblock {\em J. Chem. Phys.} {\bf 136}, 094507 (2012).

\bibitem{Holten2012b}
{\sc V.~Holten}, {\sc J.~Kalov\'{a}}, {\sc M.~A. Anisimov}, and {\sc J.~V.
  Sengers},
\newblock {\em Int. J. Thermophys.} {\bf 33}, 758 (2012).

\end{thebibliography}

%%%%%%%%%%%%%%%%%%%%%%%%%%%%%%%%%%%%%%%%%%%%%%%%%%%%%%%%%%%%%%%%%%%%%%%%%%%%%%%
\end{document}